\documentclass[
reprint, 
tightenlines, 
superscriptaddress, 
showpacs,
nobibnotes,
 amsmath,amssymb,
 aps,
pra, 
floatfix,
dvipdfmx
]{revtex4-1}

\usepackage{latexsym} 
\usepackage{enumerate}
\usepackage{bm}
\usepackage{graphicx}
\usepackage[usenames]{color} 
\usepackage{dcolumn}
\usepackage{ulem} 
\usepackage{multirow} 
\usepackage{booktabs}

\allowdisplaybreaks

\begin{document}


\title{Fast macroscopic-superposition-state generation by coherent driving}

\author{Emi Yukawa} 
\email{emi.yukawa@riken.jp}
\affiliation{National Institute of Informatics, 2-1-2 Hitotsubashi, Chiyoda-ku, Tokyo 101-8430, Japan} 
\affiliation{RIKEN Center for Emergent Matter Science, 2-1, Hirosawa, Wako-shi, Saitama, 351-0198, Japan} 
\author{G. J.  Milburn} 
\affiliation{Centre for Engineered Quantum Systems, School of Mathematics and Physics, 
The University of Queensland, St Lucia, QLD 4072, Australia} 
\author{Kae Nemoto} 
\affiliation{National Institute of Informatics, 2-1-2 Hitotsubashi, Chiyoda-ku, Tokyo 101-8430, Japan}


\date{\today}

\begin{abstract} 
We propose a scheme to generate macroscopic superposition states (MSSs) in spin ensembles, where a coherent driving field is applied to accelerate the generation of macroscopic superposition states. The numerical calculation demonstrates that this approach allows us to generate a superposition of two classically distinct states of the spin ensemble with a high fidelity above 0.97 for 300 spins. For a larger spin ensemble, though the fidelity slightly declines, it maintains above 0.84 for an ensemble of 500 spins. The time to generate an MSS is also estimated, which shows that the significantly shortened generation time allows us to achieve such MSSs within a typical coherence time of the system.  
\end{abstract}

\maketitle

\section{Introduction}
For a long time quantum mechanics has been considered as the theory to describe physical behaviour in the microscopic scale, and the quantum theory has provided the framework for the development of the technologies, which clearly characterise the twenties century. 
Semiconductor-based computer technology and laser are typical examples which require quantum-mechanical understanding in the underlying physics. 
Our effort to manipulate quantum coherence did not however stop there, and in the recent years it has continued to realize a longer coherence time and a higher fidelity. 
As one of the consequences of this development, we began to manipulate quantum coherence in macroscopic states of matter~\cite{Knee}. 

To further penetrate this new quantum regime, it is necessary, however, to circumvent experimental obstacles for a system to behave quantum mechanically in an even large scale. 
For instance, the non-classical generation of states, such as squeezed states~\cite{Yuen} and the $N00N$ states~\cite{Lee}, has its limitation in reality: squeezing becomes too noisy when squeezing gets too large, and the success probability or the fidelity of $N00N$ states is plummeted when $N$ gets larger. 
Superposition states of two or several coherent states progressively become difficult to generate as the coherent states approach to be orthogonal. 
These non-classical states are not only interesting as a promising candidate for quantum technology such as high precision measurements, but the macroscopic non-classical states are also a route to novel quantum phenomena never achievable before. 
To realise these states, as the attainable precision has its own limitation even with the best technology, it is essential to introduce a new mechanism for quantum properties to win over its decoherence. 
In this paper, we focus on collective spin systems and show how such macroscopic non-classical states can be generated.  

Collective spin states have been investigated in cold atom systems such as Bose-Einstein condensates and solid-state systems, where spins are abundant and its inhomogeneous broadening is well suppressed. 
When a state forms a superposition of two or more macroscopically distinguishable states, such as large coherent states, it is called macroscopic superposition states (MSSs)~\cite{Gerry1}. 
They are also known as $N$-particle Greenberger-Horne-Zeilinger (GHZ) states~\cite{Mermin,Bollinger1}, $N00N$ states~\cite{Lee}, or macroscopic quantum superposition states~\cite{Molmer,Milburn,Rao}, depending on what macroscopic nature we are interested in. 
These states are not only interesting for their macroscopic quantum behaviour, but they are also potentially applicable to Heisenberg-limited spectroscopy~\cite{Bollinger1,Huelga,Boto,Mitchell,Leibfried,Giovannetti,Pezze,Jones,MaConnell}, quantum computation with coherent states~\cite{Ralph,Barrett,Marek}, and quantum repeaters~\cite{Duan}, as we see them playing the central role in the implementation of quantum technology. 
Our primary interest in this Letter is a superposition state of two macroscopically distinguishable spin coherent state~\cite{Arecchi}, which we refer to as a spin cat state. 

In an ensemble of $N$ identical $1/2$-spins, a spin cat state can be generated from a separable coherent spin state (CSS)~\cite{Arecchi} via a number of ways. 
A quadratic interaction between spins~\cite{Agarwal,Molmer,Milburn,Chumakov,Pezze,Rao,Voje,Opatrny,MaConnell,Dooley1,Lau}, the QND interaction~\cite{Gerry2,Recamier}, and the dispersive Tavis-Cummings interaction~\cite{Bennett,Dooley2} generate these spin cat states, whereas a series of controlled-NOT gates~\cite{Huelga,Jones,Gao}, or a sequence of spin measurements~\cite{Kok,Nielsen,Chen} have been proposed. 
The quadratic interaction, essentially equivalent to the sequence of the controlled-NOT gates~\cite{Khaneja,Zhang}, shows better scalability with respect to the number of spins. 
This interaction is often called the one-axis twisting interaction and is given by  ${\hat{H}}_{\mathrm{OAT}} = \hbar \chi {\hat{J}}_z^2$, 
where $\chi$ represents the interaction energy and the collective spin operator is defined as ${\hat{J}}_{\mu} \equiv \frac{1}{2} \sum_{j=1}^N {\hat{\sigma}}_{\mu}^{(j)}$ ($\mu = x, y, z$) with the Pauli operator ${\hat{\sigma}}_{\mu}^{(j)}$ of the $j$th spin~\cite{Kitagawa}. 
The Hamiltonian ${\hat{H}}_{\mathrm{OAT}}$ has been implemented in ultracold $^{87}$Rb atomic gases and trapped $^9$Be$^+$ ions with $N \sim O (10^{2-4})$ spins to create squeezed spin states~\cite{Gross,Leroux,Riedel,Strobel,Bohnet}. 

Spin cat states have been experimentally created in two-level systems of trapped ions~\cite{Sackett}, high-symmetry molecules in NMR~\cite{Jones}, and circularly polarized light~\cite{Gao}. 
These cat stats are comprised of $4$-$14$ spins and do not scale up to larger spin ensembles. 
One of the main difficulties is that the cat-state preparation via the one-axis twisting interaction ${\hat{H}}_{\mathrm{OAT}}$ requires an evolution time of $t =\pi /2\chi$~\cite{Agarwal,Chumakov,Rao,Milburn,Molmer,Pezze,Dooley1} which is comparable to at best or longer than the coherence time of the spin ensemble~\cite{Li,Leroux,Gross,Riedel,Strobel,Bohnet} for the number of spins larger than $N \sim O (10^2)$. 
To create a macroscopic spin cat state, one has to maintain its coherence beyond this interaction time, which remains challenging. 

One strategy to shorten the evolution time to create the cat state is to utilize the transverse magnetic field~\cite{Cirac,Gordon,Micheli}, that is, 
${\hat{H}}_{\mathrm{LMG}} = \hbar (\chi {\hat{J}}_z^2 + \Omega {\hat{J}}_x)$. 
This Hamiltonian has been known as the Lipkin-Meshkov-Glick Hamiltonian~\cite{Lipkin} and implemented in a cold-atomic system to generate squeezed spin states~\cite{Strobel,Muessel}. 
Not only squeezed spin states but spin cat states can be expected to be created via ${\hat{H}}_{\mathrm{LMG}}$ within the evolution time of 
$t \sim O (\log {N}/\chi N)$; however, the fidelity to the cat state degrades to be $0.4 \sim 0.6$ as the number of spins increases to $\sim O (10^2)$~\cite{Micheli}.  

We here propose a scheme to apply a coherent driving field to the spin ensemble in order to speed up the cat-state creation via ${\hat{H}}_{\mathrm{OAT}}$. 
We numerically demonstrate that this scheme can generate a macroscopic superposition state with the fidelity to the ideal cat state above $0.84$ for the number of spins up to $500$. 
The time scale to generate a cat state can be made shorter than or comparable to the coherence time of atomic gases. 

\begin{figure*} 
	\begin{center} 
		\centering \includegraphics[bb = 0 0 1106 367, clip, scale = 0.44]{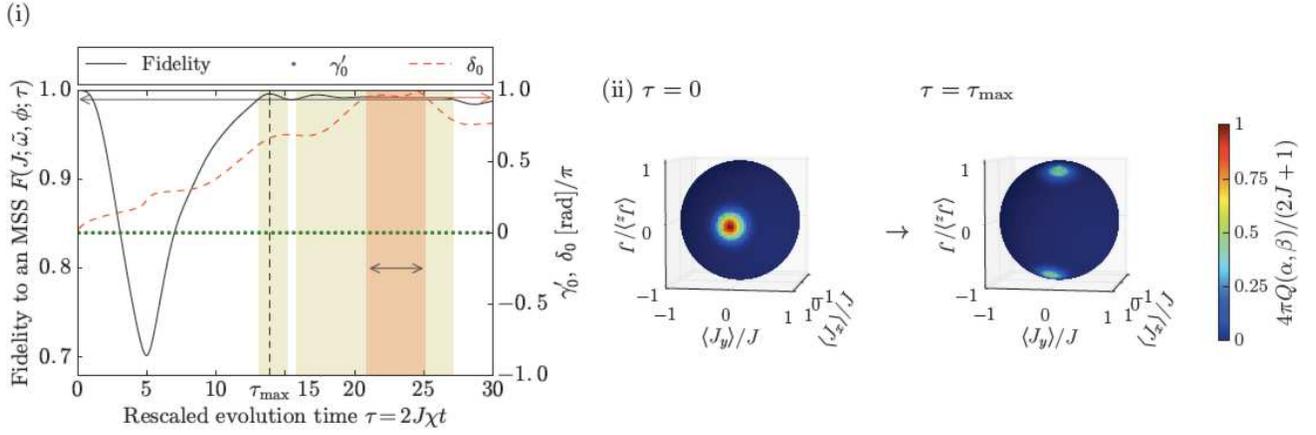}
		\caption{(Color Online) (i) Plots of time evolution of the fidelity $F(J; \tilde{\omega} ,\phi ;\tau )$, the relative phase ${\gamma}_0^{\prime}$, and the displacement angle ${\delta}_0$ and (ii) the Q-functions $Q(\alpha ,\beta ) \equiv \frac{2J+1}{4\pi} |\langle {\Phi}_{\mathrm{CSS}} (J; \alpha ,\beta )| \Psi (J; \tilde{\omega} , \phi ;\tau ) \rangle |^2$ corresponding to the initial state and the first local maximum of the fidelity for $J=50$, $\tilde{\omega} = 0.0204 \pi$ and $\phi = 0.024\pi$. (i) Time dependences of $F(J; \tilde{\omega} ,\phi ;\tau )$, ${\gamma}_0^{\prime}$, and ${\delta}_0$ are indicated by the black solid curve, the red dashed curve, and the green dots, respectively. The yellow and red shaded regions represent the intervals $F(J;\tilde{\omega}, \phi ;\tau ) \geq 0.99$ and ${\delta}_0 \geq 0.95\pi$, respectively. (ii) The color at the point indicated by the polar and azimuthal angles of $(\alpha ,\beta )$ represents $\frac{4\pi}{2J+1} Q(\alpha ,\beta )$ according to the right gauge. The time evolution of the fidelity and the Q-functions for $J=74.5$ and $J=200$ are shown in Figs.~\ref{fig:timedependence} and \ref{fig:QfuncMSS}, respectively. } 
		\label{fig:fidelityN101}
	\end{center}
\end{figure*} 
\begin{figure*} 
	\begin{center} 
		\centering \includegraphics[bb = 0 0 885 780, clip, scale = 0.44]{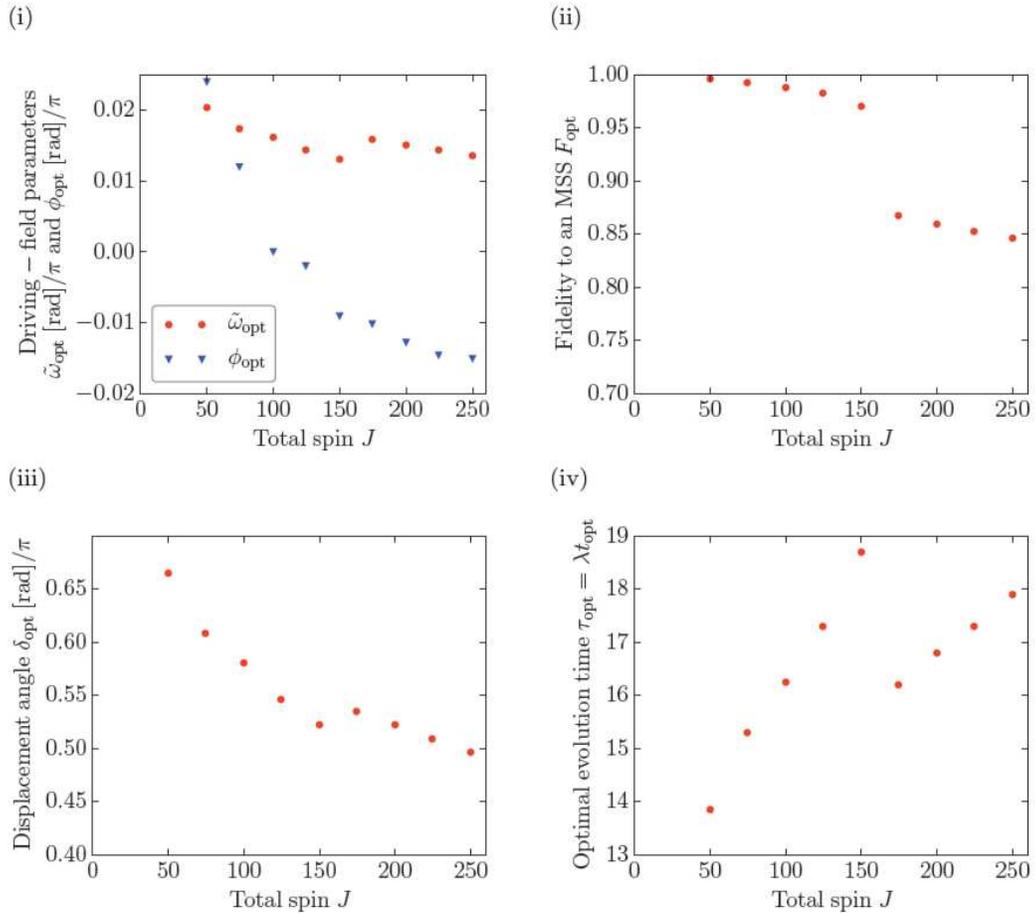}
		\caption{Total spin dependences of (i) the set of the rescaled driving frequency and the driving phase, (ii) the fidelity, (iii) the displacement angle, and 
		(iv) the rescaled evolution time. In the plots (i)-(iv), there are discontinuity between $J=150$ and $174.5$. }
		\label{fig:Fcat-vs-J}
	\end{center}
\end{figure*} 
\section{Model and Method}
We consider a collective spin system consisting of $N$ identical $1/2$ spins with two degrees of freedom $| \uparrow \rangle$ and $| \downarrow \rangle$. 
A single-spin state can be parametrized as $|\alpha ,\beta \rangle \equiv \cos \frac{\beta}{2} | \uparrow \rangle + e^{i\alpha} \sin \frac{\beta}{2} | \downarrow \rangle$ 
in terms of the polar and azimuth angles $(\alpha ,\beta)$ ($\alpha \in (-\pi ,\pi ]$ and $\beta \in [0, \pi ]$). 
A CSS of the $N$-spin ensemble can also be expressed in terms of $\alpha$ and $\beta$ as $| {\Phi}_{\mathrm{CSS}} (J; \alpha ,\beta ) \rangle = |\alpha ,\beta {\rangle}^{\otimes 2J} = \sum_{n=0}^{2J} \sqrt{_{2J}{\mathrm{C}}_n} \ {\cos}^{2J-n} \frac{\beta}{2} \ {\sin}^n \frac{\beta}{2} \ e^{in\alpha } |J, J-n \rangle$, where $J = N/2$ represents the total spin, $_m{\mathrm{C}}_n$ represents the number of $n$ combinations out of $m$ elements, and $|J, M\rangle $ denotes the eigenstate of the collective spin operator ${\hat{J}}_z$ corresponding to the eigenvalue $M$. 
Setting $| {\Phi}_{\mathrm{CSS}} (J; 0 ,\frac{\pi}{2} ) \rangle $ as the initial state, we consider the time evolution by the Hamiltonian composed of the one-axis twisting 
Hamiltonian and the coherent driving field,
\begin{equation} 
	\hat{H} (J; t) = \hbar [ \chi {\hat{J}}_z^2 + \Omega {\hat{J}}_x \cos {(\omega t + \phi)}], \label{eq:H}
\end{equation} 
where $\Omega$, $\omega$, and $\phi$ denote the driving energy, the driving frequency, and the phase of the driving 
field, respectively. 
Here, we define $\lambda \equiv 2 \chi J$ and rescale the elapsed time, the driving energy, and the driving frequency as $\tau \equiv \lambda t$, $r \equiv \Omega / \lambda$, 
and $\tilde{\omega}\equiv \omega/\lambda $. 
Throughout the paper, $r$ is fixed at $r= 1$, while $J$, $\omega$, and $\phi $ are left to be tunable. 
Under the Hamiltonian~(\ref{eq:H}), the initial state evolves as 
\begin{equation} 
	|\Psi (J; \tilde{\omega} , \phi  ;\tau ) \rangle \equiv e^{- i \int_{{\tau}^{\prime} = 0}^{\tau} d{\tau}^{\prime} \tilde{h} ({\tau}^{\prime} )} \ | {\Phi}_{\mathrm{CSS}} (J; 0 ,\pi /2) \rangle ,  
\end{equation} 
where $\tilde{h} (\tau ) \equiv \frac{1}{2J} {\hat{J}}_z^2 + {\hat{J}}_x \cos {(\tilde{\omega}\tau + \phi )}$. 
When $\tilde{\omega} $ is moderately slow and $\phi\simeq 0$, we can expect the initial $x$-polarized CSS to become a superposition of two CSSs, via the highest-energy eigenstate transfer and the preservation of the relative phase ${\gamma}_M^{\prime}$ between $|J, \pm M \rangle$ by the time-dependent Hamiltonian in Eq.~(\ref{eq:H}). 
The initial state is close to the highest energy eigenstate of the Hamiltonian Eq.~(\ref{eq:H}) for a small $\phi$ at $\tau = 0$ and the initial state evolves, following the highest energy eigenstate of Eq.~(\ref{eq:H}), which ends up to be a superposition of two coherent spin states at a certain $\tau$ satisfying $0 < \omega \tau + \phi \lesssim \pi / 2$.
Although the gap between the highest and the second highest energy eigenstates closes during the process, the relative phases ${\gamma}_M^{\prime}$'s are robust against the breakdown of the adiabatic condition for the time-dependent Hamiltonian. 
This is because Eq.~(\ref{eq:H}) is preserves ${\gamma}_M^{\prime}$'s, and $^{\forall} {\gamma}_M^{\prime} = 0$ for the highest energy eigenstate and the initial state, whereas $^{\forall} {\gamma}_M^{\prime} = \pi$ for the second highest energy eigenstate, as detailed in Appendix~\ref{sec:s1}.  

An MSS can be parametrized in terms of the superposition phase $\gamma$ in addition to $\alpha$ and $\beta$ characterizing a CSS as in Ref.~\cite{Rao}: 
\begin{widetext}
\begin{align} 
	| {\Phi}_{\mathrm{MSS}} (J; \alpha ,\beta ,\gamma ) \rangle 
	&\equiv \frac{1}{A(J; \alpha ,\beta ,\gamma )} \left (| {\Phi}_{\mathrm{CSS}} (J; \alpha ,\beta ) \rangle 
	+ e^{i\gamma} | {\Phi}_{\mathrm{CSS}} (J; - \alpha ,\pi - \beta ) \rangle \right ) \nonumber \\
	&= \frac{1}{A(J; \alpha ,\beta ,\gamma )} \sum_{n=0}^{2J} \sqrt{_{2J}{\mathrm{C}}_n} \ {\cos}^{2J-n} \frac{\beta}{2} \ {\sin}^n \frac{\beta}{2} \ e^{in\alpha} \left (
	|J,J-n \rangle + e^{i {\gamma}^{\prime}} |J, -J+n \rangle \right ), \label{eq:cat}  
\end{align} 
\end{widetext}
where $\gamma \in (-\pi ,\pi]$ and $(\alpha ,\beta ,\gamma) \neq (0, \frac{\pi}{2}, \gamma)$, $(\pi ,\frac{\pi}{2}, \gamma )$. 
Here, the normalisation constant $A(J; \alpha ,\beta ,\gamma )$ is defined as 
\begin{align} 
	A(J; \alpha ,\beta ,\gamma ) &\equiv \sqrt{2[1 +  {\cos}^{2J} \alpha \ {\sin}^{2J} \beta \ \cos {(\gamma - 2J\alpha )}]} \nonumber \\ 
	&= \sqrt{2[1 + {\cos}^{2J} \alpha \ {\sin}^{2J} \beta \ \cos {{\gamma}^{\prime}}]}. \label{eq:norm}
\end{align}
In the second expression above, we introduce a new relative phase ${\gamma}^{\prime} \equiv \gamma - 2J\alpha$ (${\gamma}^{\prime} \in (-\pi ,\pi]$) between two ${\hat{J}}_z$ eigenstates $|J, M\rangle$ and $|J, -M\rangle$ to characterize the MSS, since this relative phase is the parameter relevant to interferometry as shown later and detailed in Appendix~\ref{sec:s3}. 
The displacement angle $\delta$ between two superposed CSSs can be expressed in terms of $\alpha$ and $\beta$ as  
\begin{align} 
	&\delta (\alpha ,\beta ) \nonumber \\ 
	=& \pi - \arccos {\left \{ \frac{1}{2} [1-\cos {2\alpha} + (1+\cos {2\alpha} ) \cos {2\beta} ] \right \} }.
\end{align}    
The fidelity of the state $|\Psi (J; \tilde{\omega} , \phi ;\tau ) \rangle$ to the MSS in Eq.~(\ref{eq:cat}) is obtained by
\begin{align}   
	&F (J; \tilde{\omega} , \phi ; \tau ) \nonumber \\ 
	\equiv & \max_{\alpha ,\beta ,{\gamma}^{\prime}} [ |\langle {\Phi}_{\mathrm{MSS}} (J; \alpha ,\beta ,{\gamma}^{\prime}) | \Psi (J; \tilde{\omega} , \phi ; \tau ) \rangle |^2 ], \label{eq:fid} \\ 
	& ({\alpha}_0 (J; \tilde{\omega} , \phi ; \tau ) ,{\beta}_0 (J; \tilde{\omega} , \phi ; \tau ) ,{\gamma}^{\prime}_0 (J; \tilde{\omega} , \phi ; \tau ) ) \nonumber \\ 
	 \equiv & \underset{\alpha ,\beta ,{\gamma}^{\prime}}{\mathrm{argmax}} \ [ |\langle {\Phi}_{\mathrm{MSS}} (J; \alpha ,\beta ,{\gamma}^{\prime}) | \Psi (J; \tilde{\omega} , \phi ; \tau ) \rangle |^2 ], 
\end{align} 
where $F (J; \tilde{\omega} , \phi ; \tau )$ is numerically maximised with respect to $\alpha$, $\beta$, and ${\gamma}^{\prime}$ by the basin-hopping method~\cite{basin,Wales}. 
The fidelity in $F (J; \tilde{\omega} , \phi ; \tau )$ in Eq.~(\ref{eq:fid}) for fixed $\tilde{\omega}$ and $\phi$ has a local maximum at the rescaled elapsed time $\tau = {\tau}_{\mathrm{max}}$ as shown in Fig.~\ref{fig:fidelityN101} (i). 
At ${\tau}_{\mathrm{max}}$, the Q-function becomes a superposition of two CSSs as shown in Figs.~\ref{fig:fidelityN101} (ii). 
We numerically obtain ${\tau}_{\mathrm{max}}$ and the fidelity of the first local maximum $F (J; \tilde{\omega} , \phi ; {\tau}_{\mathrm{max}} ) \equiv F_{\mathrm{max}} (J; \tilde{\omega} , \phi)$. 
After $\tau = {\tau}_{\mathrm{max}} (J; \tilde{\omega} ,\phi )$, the state maintains high fidelity for quite a while as shown in Fig.~\ref{fig:fidelityN101} (i), which implies that the fidelity is rather insensitive to timing in creating an MSS via this method (see also Appendix~\ref{sec:s2} and Figs.~\ref{fig:timedependence}). 
We also note that ${\gamma}_0^{\prime}$ is time-independent during the time evolution given by Eq.~(\ref{eq:H}) as shown in Fig.~\ref{fig:fidelityN101} (i), which implies that the phase ${\gamma}_M^{\prime}$ between $|J, M\rangle$ and $|J, - M\rangle$ is preserved under the Hamiltonian in Eq.~(\ref{eq:H}). 

Next, in order to investigate the driving frequency and its phase optimising the fidelity and the displacement angle at ${\tau}_{\mathrm{max}}$, 
we plot $F_{\mathrm{max}} (J; \tilde{\omega}, \phi )$ and displacement angle ${\delta}_{\mathrm{max}} (J; \tilde{\omega}, \phi )$ with respect to $\tilde{\omega} $ and $\phi $ for $J=50-250$ in Appendix~\ref{sec:s2}. 
We estimate $\tilde{\omega} = {\tilde{\omega}}_{\mathrm{opt}} (J)$ and $\phi = {\phi}_{\mathrm{opt}} (J)$ maximising $F_{\mathrm{max}} (J; \tilde{\omega}, \phi )$ under the 
condition ${\delta}_{\mathrm{max}} (J; \tilde{\omega}, \phi ) > 0.4\pi$ and plot $F_{\mathrm{opt}} (J) \equiv F_{\mathrm{max}} (J; {\tilde{\omega}}_{\mathrm{max}}, {\phi}_{\mathrm{max}})$, ${\delta}_{\mathrm{opt}} (J) \equiv {\delta}_{\mathrm{max}} (J; {\tilde{\omega}}_{\mathrm{opt}}, {\phi}_{\mathrm{opt}} )$, ${\tilde{\omega}}_{\mathrm{opt}} (J)$, ${\phi}_{\mathrm{opt}} (J)$, and ${\tau}_{\mathrm{opt}} (J) \equiv {\tau}_{\mathrm{max}} (J; {\tilde{\omega}}_{\mathrm{opt}}, {\phi}_{\mathrm{opt}})$ against the total spin $J$ in Figs.~\ref{fig:Fcat-vs-J}. 
The maximum fidelity jumps in the regime $150 \leq J \leq 174.5$, which is caused by a finite probability distribution around $\alpha =0$ and $\beta = \pi /2$ at ${\tau}_{\mathrm{max}}$ as shown in the Q-functions in Figs.~\ref{fig:QfuncMSS} of Appendix~\ref{sec:s2}. 

\begin{figure*}
	\begin{center}
		\centering \includegraphics[bb = 0 0 909 771, clip, scale = 0.44]{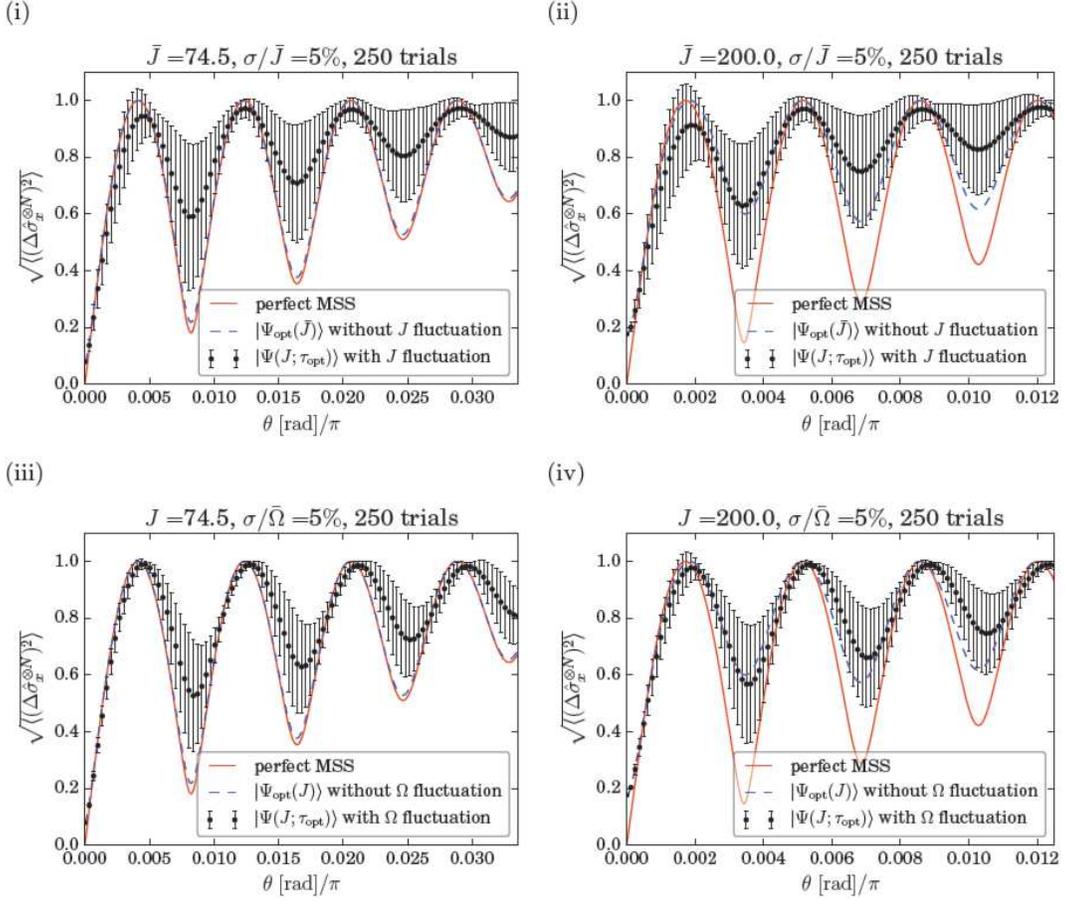}
		\caption{(Color Online) Interference fringes of the quantum fluctuations in ${\hat{\sigma}}_x^{\otimes N}$. The red solid curves and the blue dashed curves indicate the interference fringes produced by the perfect MSS given in Eq.~(\ref{eq:fringe}) and pure MSSs $|{\Psi}_{\mathrm{opt}} (J) \rangle$, respectively. 
		The black dots with error bars represent the mean values and the standard variances of the quantum fluctuations in ${\hat{\sigma}}_x^{\otimes N}$ and ${\tau}_{\mathrm{opt}}$ of $|\Psi (J; {\tau}_{\mathrm{opt}} ) \rangle$ represents the optimised evolution time for $\bar{J}$. 
		(i) (ii) The fringes produced by the MSSs with $5\%$ of the Gaussian fluctuations in the number of spins for $N=149$ and $200$. (iii) (iv) The fringes produced by the MSSs with $5\%$ of the uniform distribution in the magnitude of the driving field $\Omega$.}
		\label{fig:fringes}
	\end{center} 
\end{figure*}
\begin{figure*}
	\begin{center}
		\centering \includegraphics[bb = 0 0 923 407, clip, scale = 0.44]{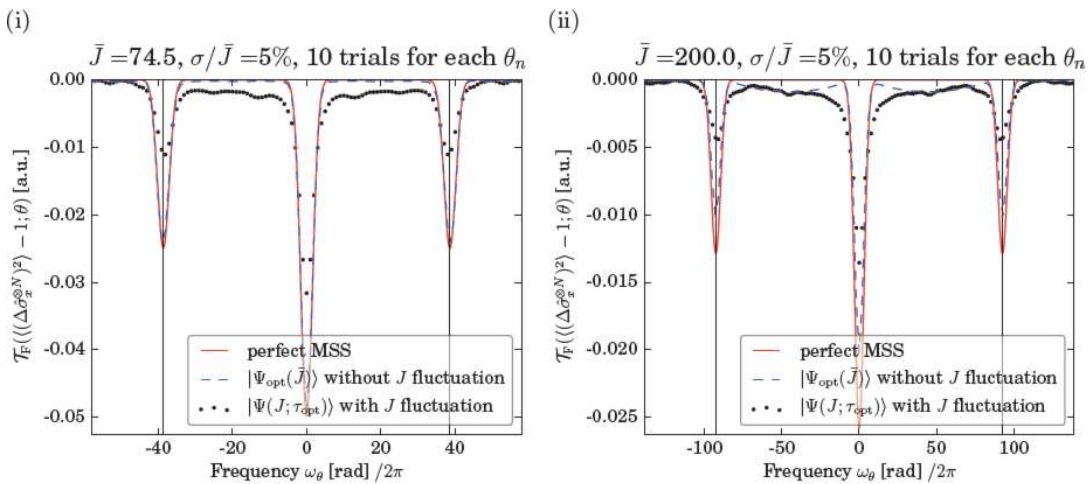}
		\caption{(Color Online) Discrete Fourier transformation of ${\hat{\sigma}}_x^{\otimes N} - 1$. The red solid curves and the blue dashed curves indicate the spectra produced by the perfect MSS given in Eq.~(\ref{eq:fringe}) and pure MSSs $|{\Psi}_{\mathrm{opt}} (J) \rangle$. 
		The black dots represent the mean values of the spectra and ${\tau}_{\mathrm{opt}}$ of $|\Psi (J; {\tau}_{\mathrm{opt}} ) \rangle$ represents the optimised evolution time for $\bar{J}$. 
		The black solid lines indicate the frequencies of the interference fringes for the perfect MSS, i.e., $\omega = \pm 4J \cos {\beta}$ in Eq.~(\ref{eq:fringe}). 
		(i) (ii) The spectra produced by the MSSs with $5\%$ of the Gaussian fluctuations in the number of spins for $N=149$ and $200$ at each step of rotation.}
		\label{fig:fft}
	\end{center} 
\end{figure*}

\section{Nonclassicality Witness and Precision Measurements}
To witness the nonclassicality of the generated MSS ${|\Psi}_{\mathrm{opt}} (J) \rangle \equiv |\Psi (J; {\tilde{\omega}}_{\mathrm{opt}}, {\phi}_{\mathrm{opt}} ;{\tau}_{\mathrm{opt}} ) \rangle$ in experiments, we measure the parity of the spins in the $x$ direction, ${\hat{\sigma}}_x^{\otimes N}$ after rotating ${|\Psi}_{\mathrm{opt}} (J) \rangle$ along the $z$ axis by a small angle $\theta$, which is the same protocol as the Heisenberg-limited measurement using maximally entangled states~\cite{Bollinger1}. 
If the state ${|\Psi}_{\mathrm{opt}} (J) \rangle$ is a perfect MSS, i.e., $|{\Phi}_{\mathrm{MSS}} (J;{\alpha}_{\mathrm{opt}} ,{\beta}_{\mathrm{opt}}, {\gamma}^{\prime}_{\mathrm{opt}} )\rangle$, the quantum fluctuation in the parity, $\langle (\Delta {\hat{\sigma}}_x^{\otimes N})^2 \rangle \equiv \langle ({\hat{\sigma}}_x^{\otimes N})^2 \rangle - {\langle {\hat{\sigma}}_x^{\otimes N} \rangle}^2$, exhibits fringes with respect to $\theta$ as 
\begin{equation} 
	\langle (\Delta {\hat{\sigma}}_x^{\otimes N})^2 \rangle = 1- e^{-2J{\theta}^2 {\sin}^2 {\beta}_{\mathrm{opt}} } {\cos}^2 {(2J\theta \cos {{\beta}_{\mathrm{opt}}} + {\gamma}^{\prime}_{\mathrm{opt}})}, \label{eq:fringe}
\end{equation} 
whose derivation is detailed in Appendix~\ref{sec:s3}. 
On the other hand, when the state is a mixed state of two CSSs, $\langle (\Delta {\hat{\sigma}}_x^{\otimes N})^2 \rangle = 1$, i.e., no fringe can be observed as shown in Appendix~\ref{sec:s3}. 

We compare the fringes produced by perfect MSSs, MSSs ${|\Psi}_{\mathrm{opt}} (J) \rangle$ without spin number fluctuations, ${|\Psi}_{\mathrm{opt}} (J) \rangle$ with Gaussian number fluctuations of $\sigma = 5\% \times N$ spins, and ${|\Psi}_{\mathrm{opt}} (J) \rangle$ with uniform fluctuations in the driving field magnitude $(1 \pm \sigma ) \Omega$, where $\sigma = 5\%$ for $N = 149$ and $N = 400$ as shown in Figs~\ref{fig:fringes}. 
In the cold-atom experiments, the number fluctuations and the fluctuations in $\Omega$ due to magnetic-field fluctuations may fluctuate respectively by $\lesssim 5\%$~\cite{Strobel} and a few percent at least, and they are the major noise sources that degrade fringe visibility, while the preparation time $\tau$, the driving frequency $\tilde{\omega}$, and the driving phase $\phi$ can be controlled precisely enough. 
We also numerically show robustness of fringes against the nonlinear interaction energy $\lambda$, which is equivalent to robustness against $\tau$, in Appendix~\ref{sec:s3}. 
Figures~\ref{fig:fringes} imply that the major noise source is the number fluctuation rather than the driving-field fluctuation; nonetheless we still can expect to observe the nonclassicality of the state even with $10\%$ fluctuations in the number of spins as shown in Figs.~\ref{fig:fringevarN} of Appendix~\ref{sec:s3}. 

The other major noise source would be the magnetic field $B_z$ in the $z$ direction. 
The magnetic field $B_z$ gives rise to a linear Zeeman term $p {\hat{J}}_z$ in the Hamiltonian in Eq.~(\ref{eq:H}), where the linear Zeeman energy $p \equiv g|{\mu}_B| B$ with the Land{\'e} $g$-factor and the Bohr magneton ${\mu}_B$. 
The term $p {\hat{J}}_z$ harms the preservation of the relative phases ${\gamma}_M^{\prime}$'s between the two ${\hat{J}}_z$ eigenstates $|J, \pm M\rangle$ during the time evolution.  
The linear Zeeman energy can be well controlled in experiments when the driving field is switched off; however, once it is turned on, it may be an experimentally challenging to cancel the linear Zeeman energy. 
The analysis of the effects of the linear Zeeman energy and its fluctuation and how it can be circumvented are left as future problems. 

In addition to these noises, to detect the interferometric characteristics, we typically measure the spin parity in the $x$ direction in the single-spin resolution. 
In such a scenario, trapped ion systems have a clear advantage over BECs. 

The states created via our method can also be applied to precision measurements of the rotation angle $\theta$ around the $z$ axis. 
Let us consider a frequency measurement of fringes given by $\langle (\Delta {\hat{\sigma}}_x^{\otimes N})^2 \rangle - 1$ in Eq.~(\ref{eq:fringe}). 
If a perfect MSS is created, the spectrum of the fringes are given by 
\begin{align} 
	&{\mathcal{T}}_{\mathrm{F}} (J; {\omega}_{\theta} ) = \frac{1}{\sqrt{2\pi}} \int_{- \infty}^{\infty} dt (\langle (\Delta {\hat{\sigma}}_x^{\otimes N})^2 \rangle - 1 ) e^{- i{\omega}_{\theta} t} \nonumber \\ 
	 =& - \frac{\sigma}{4} \left [2 e^{-\frac{{\omega}_{\theta}^2}{2{\sigma}^2}} + e^{-\frac{{\sigma}^2}{2} ({\omega}_{\theta} - {\bar{\omega}}_{\theta})^2} + e^{-\frac{{\sigma}^2}{2} ({\omega}_{\theta} + {\bar{\omega}}_{\theta})^2} \right ],   
	\label{eq:spectrum} 
\end{align} 
where ${\omega}_{\theta}$ is the frequency, the standard variation ${\sigma}^2 \equiv 1/4J{\sin}^2 {\beta}$, and the mean value ${\bar{\omega}}_{\theta} \equiv 4J\cos {\beta}$. 
In reality, however, a deterioration in the fidelity and spin-number fluctuations cannot be ignored, and they might wipe out the spectrum. 
We numerically calculate the spectra for the optimized state $|{\Psi}_{\mathrm{opt}} (J) \rangle$ without spin-number fluctuations and the state $|\Psi (J; {\tau}_{\mathrm{opt}} ) \rangle$ with Gaussian number fluctuations of $\sigma = 5\% \times N$ spins for $149$ spins and $400$ spins. 
Here, we assume that the states are rotated by ${\theta}_n = n \Delta \theta$, where $\Delta \theta = 1/ 10 {\omega}_{\theta}$ and $n = 1, 2, \cdots ,n_{\mathrm{max}}$ such that $n_{\mathrm{max}}$ is the maximum integer satisfying ${\theta}_{n_{\mathrm{max}}} \leq 10 / \sigma$. 
For each $n$, we perform ten rotation-and-measurement procedures and the number of spins varies according to the normal distribution thoughout the procedures. 
The mean values of $\langle (\Delta {\hat{\sigma}}_x^{\otimes N})^2 \rangle - 1$ are discrete-Fourier-transformed to obtain spectra, which are shown in Figs.~\ref{fig:fft}. 
The discrete Fourier transform is defined as 
\begin{equation} 
	{\mathcal{T}}_{\mathrm{F},\mathrm{discrete}} (J; {\omega}_{\theta}) 
	= \sum_{n=0}^{n_{\mathrm{max}}} (\langle (\Delta {\hat{\sigma}}_x^{\otimes N})^2 \rangle - 1) e^{-i {\omega}_{\theta} \frac{{\theta}_{n}}{N}}, \label{eq:discretefft}
\end{equation} 
which relates to Eq.~(\ref{eq:spectrum}) as 
\begin{equation} 
	{\mathcal{T}}_{\mathrm{F}} (J; {\omega}_{\theta} ) \simeq \sqrt{\frac{2}{\pi}} \ \Delta \theta \ \mathrm{Re} \ [{\mathcal{F}}_{\mathrm{discrete}} (J; {\omega}_{\theta}) ]. \label{eq:relation}
\end{equation} 
In Figs.~\ref{fig:fft}, we plot $\sqrt{2/\pi} \ \Delta \theta \ \mathrm{Re} \ [{\mathcal{F}}_{\mathrm{discrete}} (J; {\omega}_{\theta}) ]$ of the optimized states without spin-number fluctuations and the states $|\Psi (J; {\tau}_{\mathrm{opt}} ) \rangle$ with Gaussian number fluctuations of $\sigma = 5\% \times N$ spins and compare them with those of the ideal MSSs given in Eq.~(\ref{eq:spectrum}). 
When the number of spins is relatively small, i.e., $N=149$, the state is almost a perfect MSS in the case without a spin-number fluctuations, and we can expect to observe clear dips at ${\omega}_{\theta} = \pm {\bar{\omega}}_{\theta}$. 
When the number of spins increases to be $N=400$, the decrease in fidelity makes the dips shallower; however, they are still clearly seen. 
The Gaussian number fluctuations of $\sigma = 5\% \times N$ spins halve the depth of dips, while their positions remain almost unchanged, which indicates that the states created via our method can be applied to probes of precision measurements and sensing. 

\section{Discussion and Conclusion}
Finally we evaluate the time to generate a MSS state and compare the generation time with the coherence times reported in Refs.~\cite{Strobel,Bohnet}. 
First, we consider the case of the two-level system consisting of spin up and down states of $^9$Be$^+$ ions in a two-dimensional triangular lattice~\cite{Bohnet}. 
The interaction energy and the coherence time are respectively estimated to be $26[\mathrm{Hz}]$ and $11[\mathrm{ms}]$ for $\sim 130$ ions. Here, the major source of decoherence is spontaneous emission from an off-resonant laser beam creating uniform $z$-$z$ coupling between spins. 
For $149$ spins, we can estimate the generation time for an MSS to be $\sim 3.9[\mathrm{ms}]$, which is two order of magnitude shorter than that for the OAT interaction given by $\sim 120[\mathrm{ms}]$ and sufficiently smaller than the coherence time. 

Next, we consider the two-level system consisting of $|F=1, m_F=1 \rangle$ and $|F=2, m_F=-1 \rangle$ of cold $^{87}$Rb atoms~\cite{Strobel}. 
The major source of decoherence is the atom-number decay caused by the $1/e$ decay of the $|2,-1\rangle$ state, inelastic scattering, and three-body recombination. 
The interaction energy and the coherence time are respectively assumed to be $\chi \sim 0.44[\mathrm{Hz}]$ and $110[\mathrm{ms}]$ for $\sim 400$ atoms, whereas the coherence time for 500 spins can be estimated as $\sim 81$[ms].  In this case, the coherence time is comparable to the MSS evolution time, which is again two order of magnitude faster than the evolution time $t = \pi / \chi \sim 7.1[\mathrm{s}]$ to obtain an MSS via the OAT interaction. 
A stronger interaction between atoms could make a cold atom system to be a better candidate which shortens the MSS creation time. 

The speedup on the MSS generation time tends to be more prominent when the ensemble size gets larger, and it can be a significant advantage of this scheme to experimentally generate and test these states. 
These numbers are promising for relatively large spin ensembles to form a MSS with the current technology. 

\appendix 
\begin{figure*}
	\centering \includegraphics[bb = 0 0 1067 430, clip, scale = 0.44]{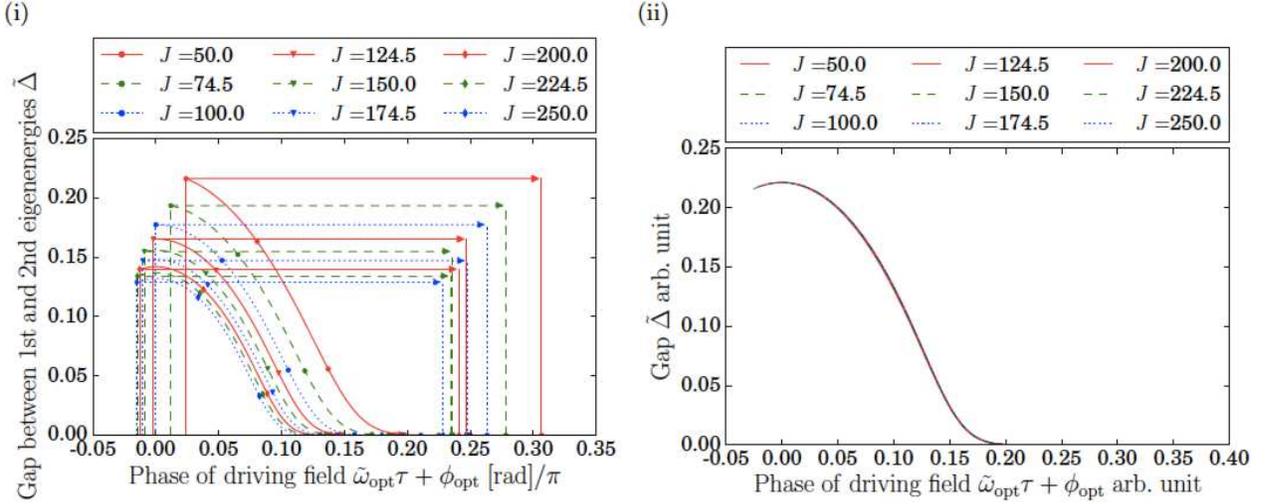}
	\caption{(Color Online) Dependence of the gap $\tilde{\Delta} (J; \tau )$ between $|{\varepsilon}_1 (J; \tau ) \rangle$ and $|{\varepsilon}_2 (J; \tau ) \rangle$ on the phase of the driving field ${\tilde{\omega}}_{\mathrm{opt}} \tau + {\phi}_{\mathrm{opt}}$. (i) The gap $\tilde{\Delta} (J; \tau )$ with respect to ${\tilde{\omega}}_{\mathrm{opt}} \tau + {\phi}_{\mathrm{opt}}$ for $J=50$-$250$. The dots, the triangles, and the thin diamonds mark $\tau = 0$, $\tau = 0.2 {\tau}_{\mathrm{opt}}$, $\tau = 0.4 {\tau}_{\mathrm{opt}}$, $\tau = 0.6 {\tau}_{\mathrm{opt}}$, $\tau = 0.8 {\tau}_{\mathrm{opt}}$, $\tau = {\tau}_{\mathrm{opt}}$ on the curves representing $\tilde{\Delta} (J; \tau )$. (ii) The gap functions $\tilde{\Delta} (J; \tau )$ for $J=50$-$250$ coincide with each other by shifts in the phase of the driving field and enlargements (or shrinks) of the magnitude of the gap energy. }
	\label{fig:gap} 
\end{figure*} 
\section*{ACKNOWLEDGMENTS} 
The authors would thank Shane Dooley, Takeshi Fukuhara, Michael Hanks, C. A. Holmes, Seth Lloyd, William J. Munro, Nguyen Thanh Phuc, Yoshiro Takahashi, and Masahito Ueda. 
This work is supported by MEXT Grant-in-Aid for Scientific Research on Innovative Areas KAKENHI Grant Number JP15H05870, MEXT Grant-in-Aid for Scientific Research(S) KAKENHI Grant Number JP25220601, and CREST, Japan Science and Technology Agency. 

\begin{figure*} 
	\centering \includegraphics[bb = 0 0 1153 1011, clip, scale = 0.44]{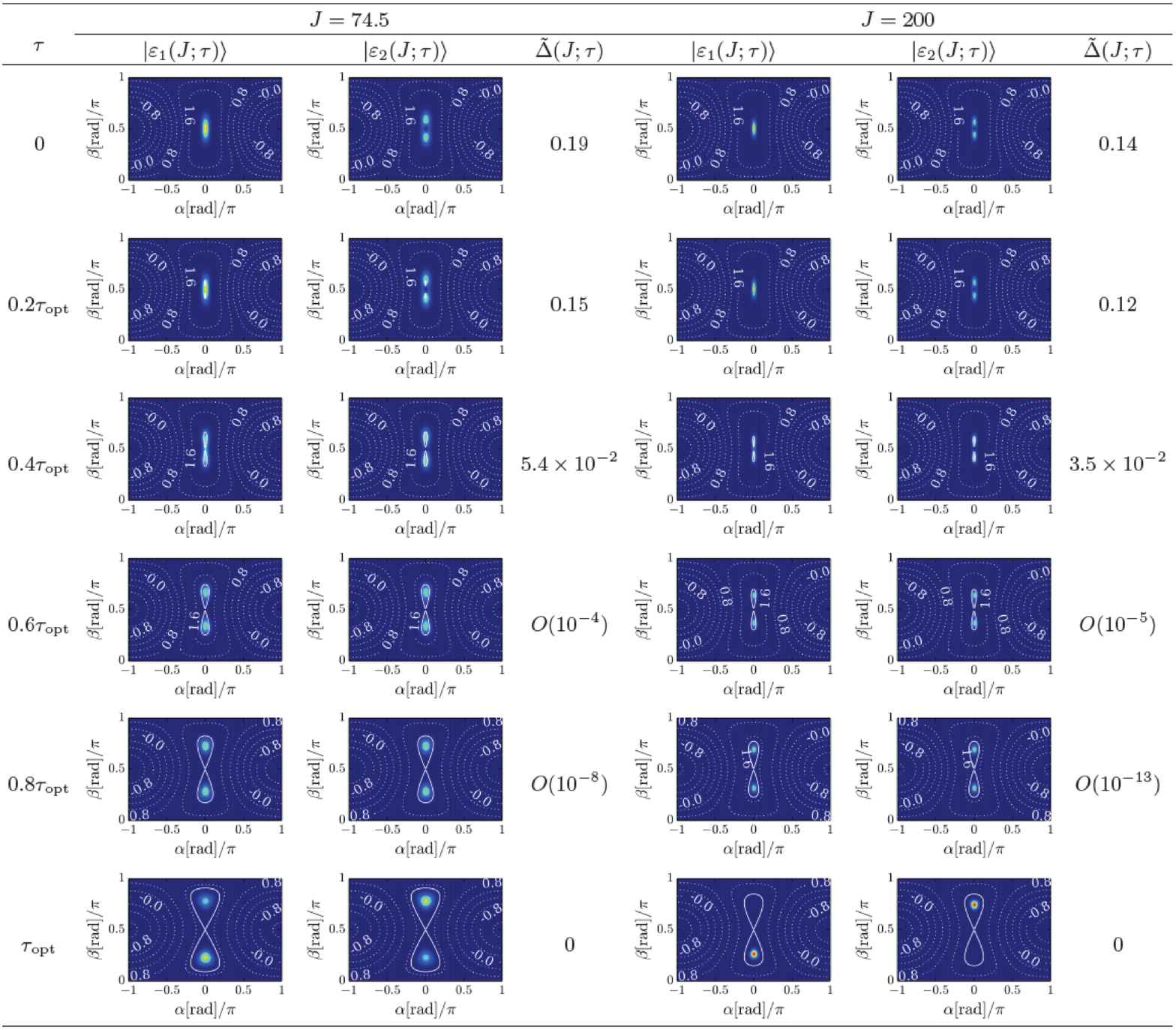}
	\caption{(Color Online) Time evolution of the Q-functions of $|{\varepsilon}_1 (J; \tau ) \rangle$ and $|{\varepsilon}_2 (J; \tau ) \rangle$ for $J=74.5$ and $J=200$. The color hue represents $\frac{4\pi}{2J+1} Q(\alpha ,\beta )$ whose gauge is the same as Figs.~1 (ii). The dotted white curves indicate the contours of ${\tilde{h}}_{\mathrm{opt}} (J; \tau )$ in the mean-field limit, which is given by ${\tilde{E}}_{\mathrm{opt}} (J; \alpha ,\beta ;\tau ) = J (\frac{1}{2} {\cos}^2 \beta + r \cos {\alpha} \sin {\beta} \cos {({\tilde{\omega}}_{\mathrm{opt}} \tau + {\phi}_{\mathrm{opt}})})$, and the values on the contours represent $\frac{2}{J} {\tilde{E}}_{\mathrm{opt}} (J; \alpha ,\beta ;\tau )$. The solid white curves represent the energy contours ${\tilde{E}}_{\mathrm{opt}} (J; \alpha ,\beta ;\tau ) = rJ \cos {({\tilde{\omega}}_{\mathrm{opt}} \tau + {\phi}_{\mathrm{opt}} )}$. } 
	\label{fig:Qfunc} 
\end{figure*} 
\begin{figure*} 
	\centering \includegraphics[bb = 0 0 445 403, clip, scale = 0.44]{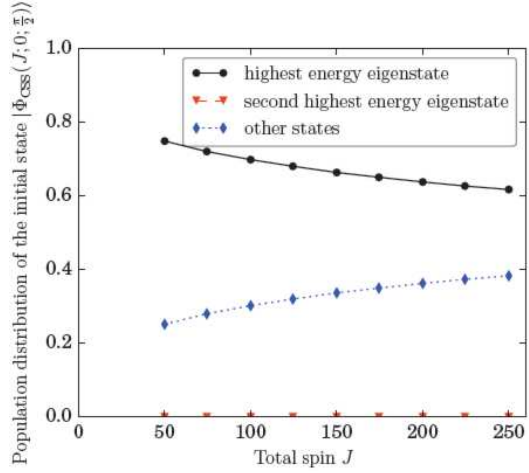}
	\caption{(Color Online) The $J$ dependence of the probability distribution of the initial state $|{\Phi}_{\mathrm{CSS}} (J; 0, \frac{\pi}{2} ) \rangle$ on the highest energy eigenstate $|{\varepsilon}_1 (J; 0) \rangle$ (black solid curve with dots) and the second-highest energy eigenstate $|{\varepsilon}_2 (J; 0) \rangle$ (red dashed curve with triangles) of the Hamiltonian $\tilde{h} (J; 0)$ and the other eigenstates (blue dotted curve with thin diamonds). The probability on $|{\varepsilon}_1 (J; 0) \rangle$ monotonically and slowly decreases with respect to $J$ and converges toward $\sim 0.5$, while the probability on $|{\varepsilon}_2 (J; 0) \rangle$ stays at $0$. The probability distributing on the other eigenstates monotonically and slowly increases and converges toward $\sim 0.5$. }
	\label{fig:population} 
\end{figure*} 
\begin{figure*} 
	\centering \includegraphics[bb = 0 0 1022 431, clip, scale = 0.44]{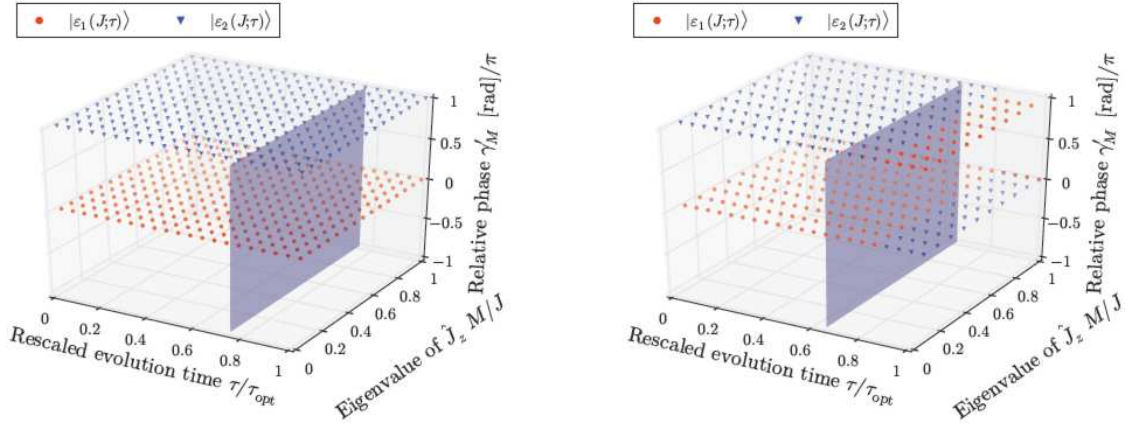} 
	\caption{(Color Online) Plots of the relative phases ${\gamma}_M^{\prime}$'s of $|{\varepsilon}_1 (J; \tau ) \rangle$ and $|{\varepsilon}_2 (J; \tau ) \rangle$ as functions of $\tau$ and $\frac{M}{J}$ for (i) $J=74.5$ and (ii) $J=200$. The red dots and the blue triangles represent ${\gamma}_M^{\prime}$'s for $|{\varepsilon}_1 (J; \tau ) \rangle$ and $|{\varepsilon}_2 (J; \tau ) \rangle$, respectively. The righthand sides of blue shaded planes parallel to the $\frac{M}{J}-{\gamma}_M^{\prime}$ planes are the time regions where $\tilde{\Delta} (J ; \tau ) < O (10^{-6})$, i.e., the region where the gaps can be regarded to be closed. We note that we plot ${\gamma}_M^{\prime}$'s for every 5 points for $J=74.5$ and for every 20 points for $J=200$ with respect to $M/J$ for the sake of visibility of the points. For both $J = 74.5$ and $200$, ${\gamma}_M^{\prime} = 0$ for $|{\varepsilon}_1 (J; \tau ) \rangle$ and ${\gamma}_M^{\prime} = \pi$ for $|{\varepsilon}_2 (J; \tau ) \rangle$ when the gaps are open. When $J=200$ and the gap closes, ${\gamma}_M^{\prime}$ can be considered to be indefinite, since the Q-functions of $|{\varepsilon}_1 (J = 200; \tau = {\tau}_{\mathrm{opt}}) \rangle$ and $|{\varepsilon}_2 (J = 200; {\tau}_{\mathrm{opt}} ) \rangle$ in Figs~\ref{fig:Qfunc} imply that they are close to coherent spin states, which are separable, and the probabilities either on $|J; \pm M\rangle $ become $\sim 0$. }
	\label{fig:relativephase} 
\end{figure*} 

\section{\label{sec:s1}Mechanism of Macroscopic-superposition-state Creation} 
We discuss the time evolution of $|\Psi (J; {\tilde{\omega}}_{\mathrm{opt}}, {\phi}_{\mathrm{opt}}; \tau ) \rangle$ by the rescaled Hamiltonian to obtain the optimum MSS $|{\Psi}_{\mathrm{opt}} (J) \rangle$ given by  
\begin{align}  
	{\tilde{h}}_{\mathrm{opt}} (J; \tau ) &\equiv \tilde{h} (J; {\tilde{\omega}}_{\mathrm{opt}} , {\phi}_{\mathrm{opt}} ;\tau ) \nonumber \\ 
	&= \frac{1}{2J} {\hat{J}}_z^2 + {\hat{J}}_x \cos {({\tilde{\omega}}_{\mathrm{opt}} \tau + {\phi}_{\mathrm{opt}} )},  
\end{align} 
from $\tau = 0$ to $\tau = {\tau}_{\mathrm{opt}}$ at which $|{\Psi}_{\mathrm{opt}} (J) \rangle$ is created. 
Here, we define the highest energy eigenstate and the second-highest energy eigenstate of ${\tilde{h}}_{\mathrm{opt}} (J; \tau )$ as $|{\varepsilon}_1 (J; \tau ) \rangle$ and $|{\varepsilon}_2 (J; \tau ) \rangle$ with the eigenenvalues ${\varepsilon}_1 (J; \tau )$ and ${\varepsilon}_2 (J; \tau )$, respectively. 
We plot the gap $\tilde{\Delta} (J; \tau ) \equiv {\varepsilon}_1 (J; \tau ) - {\varepsilon}_2 (J; \tau )$ between $|{\varepsilon}_1 (J; \tau ) \rangle$ and $|{\varepsilon}_2 (J; \tau ) \rangle$ in Figs.~\ref{fig:gap} and the Q-functions of these two eigenstates in Figs.~\ref{fig:Qfunc}. 
The gap $\tilde{\Delta} (J; \tau )$ closes at a certain $\tau$ and the two highest eigenstates ${\varepsilon}_1 (J; \tau )$ and ${\varepsilon}_2 (J; \tau )$ become states similar to two coherent spin states (CSSs) and the phase between them cannot be determined. 

The initial state follows $|{\varepsilon}_1 (J; \tau ) \rangle$ until the gap closes, since the initial state, i.e., $|{\Phi}_{\mathrm{CSS}} (J; 0, \frac{\pi}{2} ) \rangle$, has relatively high population on $|{\varepsilon}_1 (J; 0) \rangle$, whereas it does not populate on $|{\varepsilon}_2 (J; 0) \rangle$ as shown in Fig.~\ref{fig:population}. 
In such time evolution under a gap-closing Hamiltonian, in general, the state becomes a mixed state of the highest and the second highest energy eigenstates after the gap closes, because the phase between these two states cannot be determined. 
In the case of the time evolution by ${\tilde{h}}_{\mathrm{opt}} (J ;\tau )$, however, the generated state is robust against the phase uncertainty. 
The reason can be explained as follows: 
The relative phases ${\gamma}_M^{\prime}$ between $|J, \pm M\rangle$ of both the initial state and $|{\varepsilon}_1 (J; \tau ) \rangle$ before the gap closes are $0$, while ${\gamma}_M^{\prime}$'s of $|{\varepsilon}_2 (J; \tau ) \rangle$ are $\pi$ as shown in Fig.~\ref{fig:relativephase}. 
The Hamiltonian ${\tilde{h}}_{\mathrm{opt}} (J ; \tau )$ preserves ${\gamma}_M^{\prime}$ for all $\tau$, since after an infinitesimally small time evolution by $\Delta \tau$ under ${\tilde{h}}_{\mathrm{opt}} (J ; \tau )$, the phases $|J, \pm M \rangle$ of a state $\sum_{{M}^{\prime}=-J}^J {a_{M^{\prime}} |J, M^{\prime} \rangle}$ become 
\begin{widetext} 
\begin{align} 
	& \langle J, M | {\tilde{h}}_{\mathrm{opt}} (J; \tau ) \sum_{M^{\prime}=-J}^J {a_{M^{\prime}} |J, M^{\prime} \rangle} = a_M \biggl \{ 1 - i \Delta \tau \biggl \{ \frac{M^2}{2J}  + \frac{r}{2} \Bigl [ \sqrt{(J + M)(J - M + 1)} \nonumber \\ 
	& \hspace{8.95cm} + \sqrt{(J - M)(J + M + 1)} \Bigr ] \cos {({\tilde{\omega}}_{\mathrm{opt}} \tau + {\phi}_{\mathrm{opt}} )} \biggr \} \biggr \}, \\ 
	& \langle J, - M | {\tilde{h}}_{\mathrm{opt}} (J; \tau ) \sum_{M^{\prime}=-J}^J {a_{M^{\prime}} |J, M^{\prime} \rangle} = a_{-M} \biggl \{ 1 - i \Delta \tau \biggl \{ \frac{M^2}{2J}  + \frac{r}{2} \Bigl [ \sqrt{(J + M)(J - M + 1)} \nonumber \\ 
	& \hspace{9.45cm} + \sqrt{(J - M)(J + M + 1)} \Bigr ] \cos {({\tilde{\omega}}_{\mathrm{opt}} \tau + {\phi}_{\mathrm{opt}} )} \biggr \} \biggr \}, 
\end{align} 
\end{widetext} 
so the phase between $a_M$ and $a_{-M}$ is preserved. 
Therefore, after the gap closes, the state becomes the superposition state of $|{\varepsilon}_1 (J; \tau ) \rangle$ and $|{\varepsilon}_2 (J; \tau ) \rangle$ so that ${\gamma}_M^{\prime} =0$ regardless of the value of the number of spin and other parameters in the Hamiltonian in Eq.~(1), and we can expect creation of an MSS via the Hamiltonian in Eq.~(1) even though the gap between the highest and the second highest energy eigenstates closes during the time evolution. 

\begin{figure*} 
	\centering \includegraphics[bb = 0 0 918 346, clip, scale = 0.44]{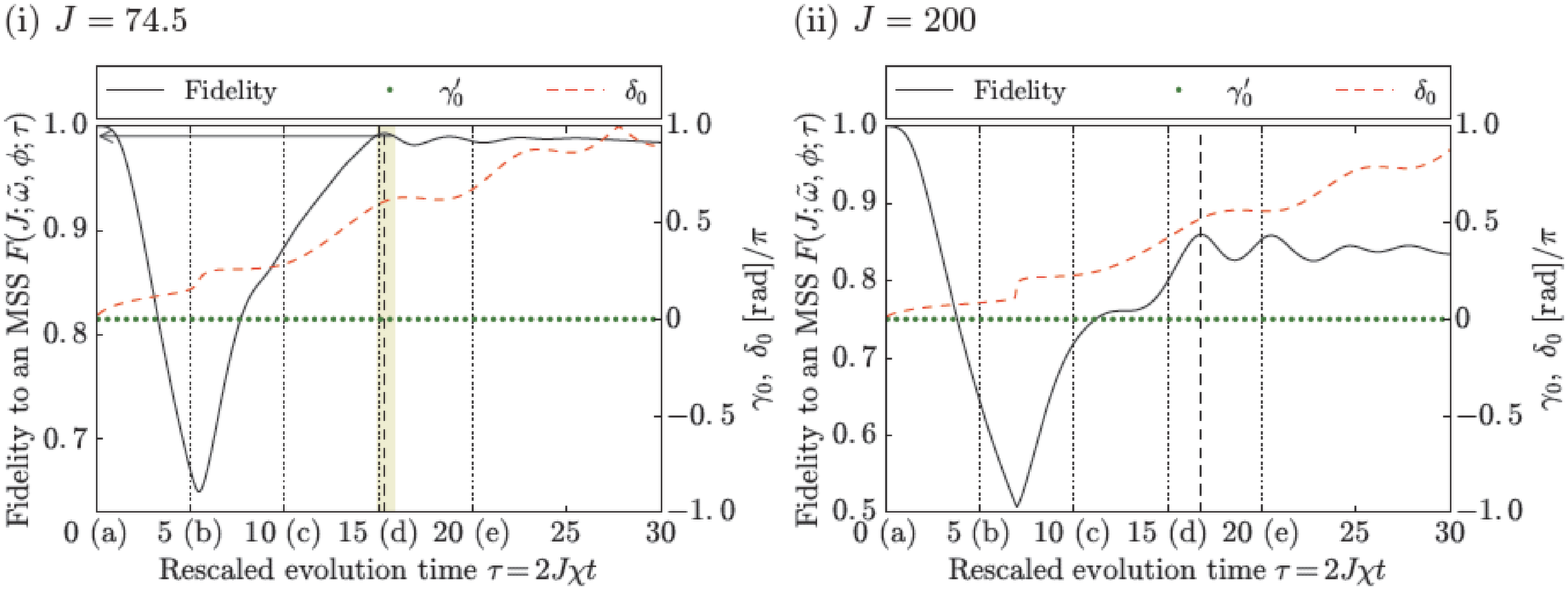}
	\caption{(Color Online) Rescaled-time dependences of fidelity $F(J; \tilde{\omega} ,\phi ;\tau )$, relative phase ${\gamma}^{\prime}_0$, and displacement angle ${\delta}_0 ({\alpha}_0 ,{\beta}_0 )$ for (i) $J=74.5$, and (ii) $J=200$. The vertical scales on the left-hand sides and the right-hand sides are for $F(J; \tilde{\omega} ,\phi ;\tau )$ and the two angles, ${\gamma}^{\prime}_0$ and ${\delta}_0 ({\alpha}_0 ,{\beta}_0 )$, respectively. The black solid curves, the green dots, and the red dashed curves represent $F(J; \tilde{\omega} ,\phi ;\tau )$, ${\gamma}^{\prime}_0$, and ${\delta}_0 ({\alpha}_0 ,{\beta}_0 )$, respectively. The yellow shaded regions and the red shaded region with a left-right arrow express the intervals satisfying $F(J; \tilde{\omega} ,\phi ;\tau ) \geq 0.99$ and the interval where an almost perfect cat state $\frac{1}{\sqrt{2}} (|J,J \rangle + |J, -J\rangle )$ is generated, i.e., the region with $F(J; \tilde{\omega} ,\phi ;\tau ) \geq 0.99$ and ${\delta}_0 ({\alpha}_0 ,{\beta}_0 ) \geq 0.95$. The Q-functions at which the black and thin dotted lines, (a)-(e), are illustrated in Figs.~\ref{fig:QfuncMSS} and $\tau = {\tau}_{\mathrm{max}}$, at which the first local maximum of the fidelity $F_{\mathrm{max}} (J; \tilde{\omega} ,\phi )$ is achieved, is indicated by the black and thin dashed line. The driving-field parameters are set to be $\tilde{\omega} = 0.0204 \pi$ and $\phi = 0.024 \pi$ for $J=50$, $\tilde{\omega} = 0.0174 \pi$ and $\phi = 0.012\pi$ for $J=74.5$, and $\tilde{\omega} = 0.0151 \pi$ and $\phi = -0.0128 \pi$ for $J=200$. } 
	\label{fig:timedependence} 
\end{figure*} 
\begin{figure*}
	\centering \includegraphics[bb = 0 0 1113 636, clip, scale = 0.44]{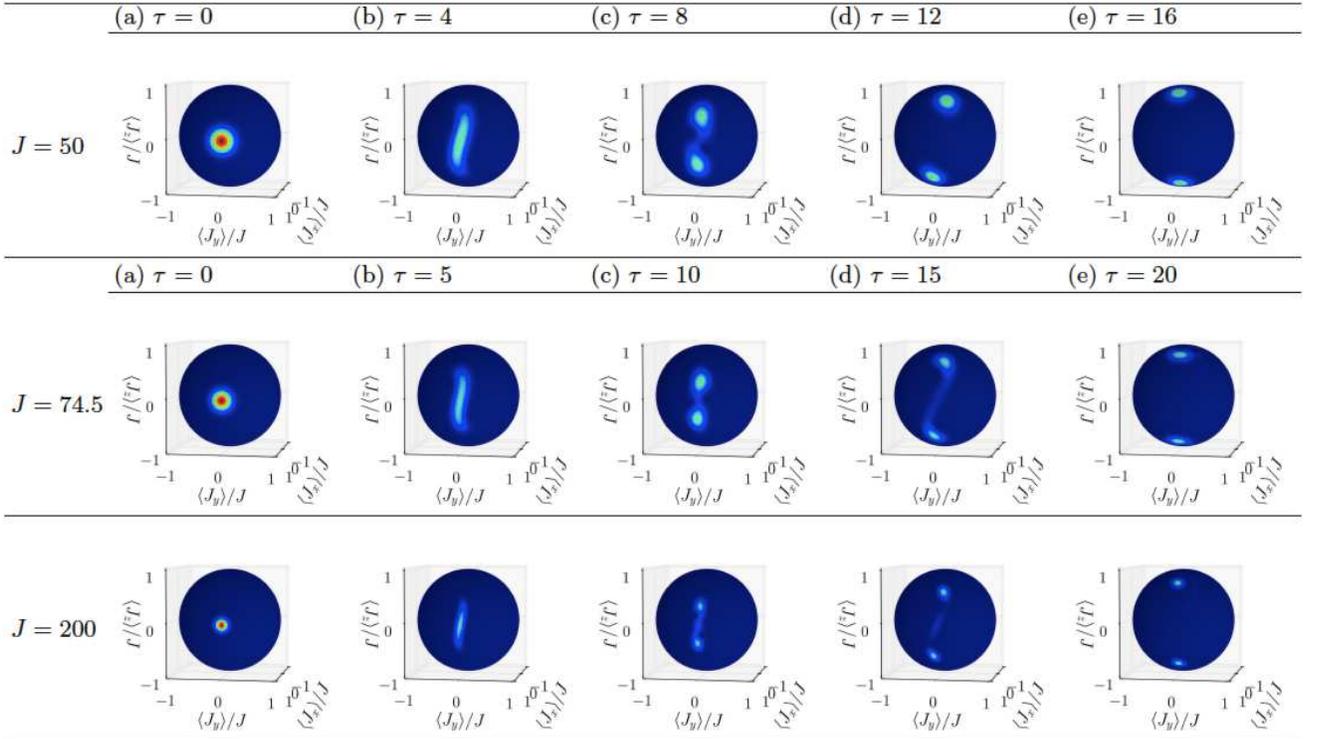}
	\caption{(Color Online) Time evolution of the Q-functions $Q(\alpha ,\beta ) \equiv \frac{2J+1}{4\pi} |\langle {\Phi}_{\mathrm{CSS}} (J; \alpha ,\beta )| \Psi (J; \tilde{\omega} , \phi ;\tau ) \rangle |^2$ for $J=50$, $J=74.5$, and $J=200$. The driving-field parameters are set to be the same as Figs.~\ref{fig:timedependence}. The color at the point indicated by the polar and azimuthal angles of $(\alpha ,\beta )$ represents $\frac{4\pi}{2J+1} Q(\alpha ,\beta )$ according to the gauge shown in Fig.~1. 
		For $J=50$, the driving field parameters are set to be $\tilde{\omega}= 0.0204\pi$ and $\phi= 0.024\pi$ and the snapshots are taken at the rescaled elapsed times of (a) $\tau =0$, (b) $4$, (c) $8$, (d) $12$, and (d) $16$. For $J=74.5$ and $J=200$, the driving field parameters are given by $\tilde{\omega}= 0.0174\pi$ and $\phi= -0.012\pi$ and $\tilde{\omega}= 0.0151\pi$ and $\phi= -0.0128\pi$, respectively, and the snapshots are taken at (a) $\tau =0$, (b) $5$, (c) $10$, (d) $15$, and (d) $20$.}
	\label{fig:QfuncMSS}
\end{figure*} 
\begin{figure*} 
	\centering \includegraphics[bb =0 0 1118 923, clip, scale = 0.44]{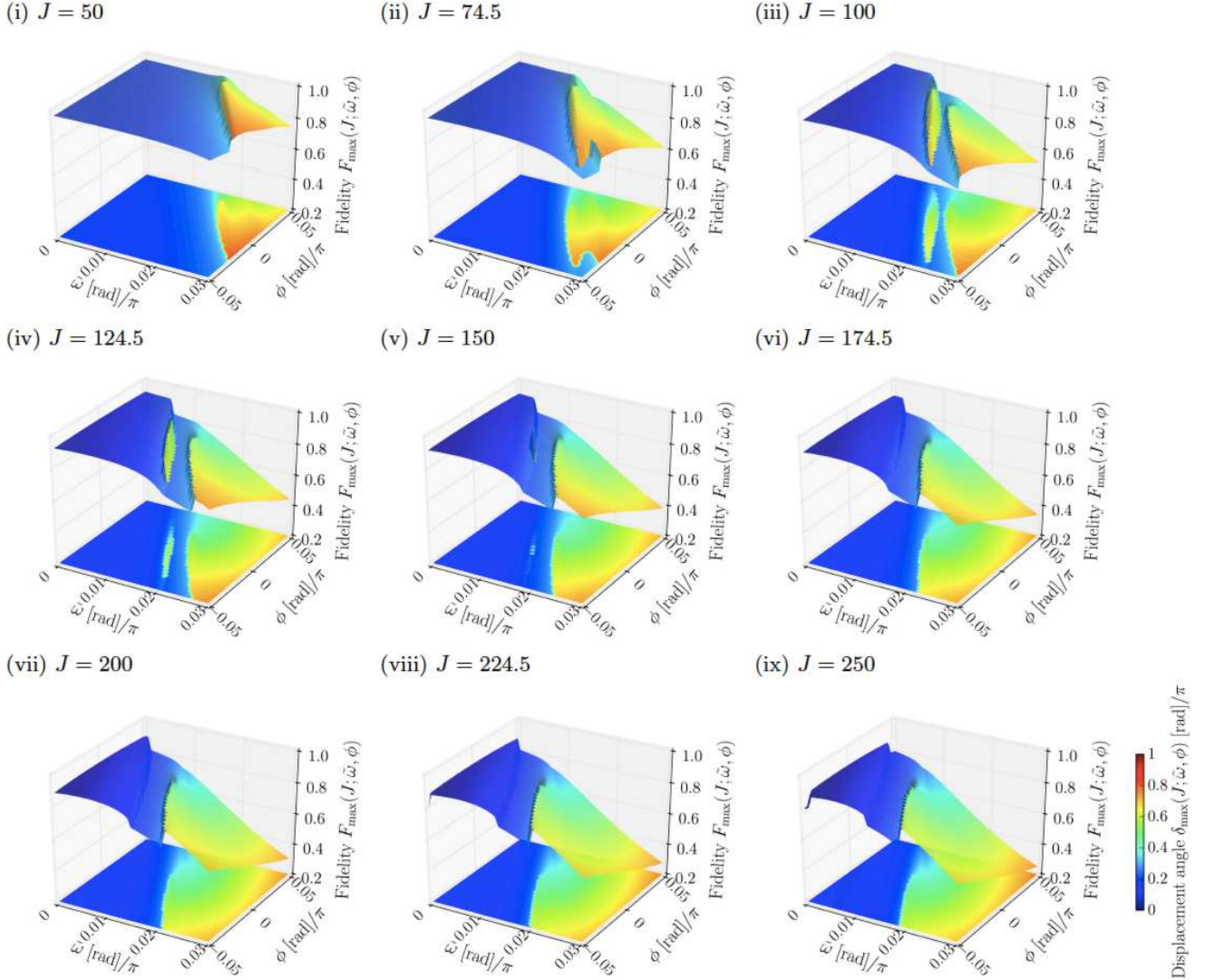}
	\caption{(Color Online) Fidelity $F_{\mathrm{max}} (J; \tilde{\omega} , \phi )$ of the first local maximum and its corresponding displacement angle ${\delta}_{\mathrm{max}} (J; \tilde{\omega}, \phi )$ as functions of the driving-field frequency and phase, $\tilde{\omega}$ and $\phi$, for (i) $J=50$, (ii) $J=74.5$, (iii) $J=100$, (iv) $J=124.5$, (v) $J=150$, (vi) $J=174.5$, (vii) $J=200$, (viii) $J=224.5$, (ix) $J=250$. The $z$ axis represent $F_{\mathrm{max}} (J; \tilde{\omega} , \phi )$ and the color on the $F_{\mathrm{max}} (J; \tilde{\omega} , \phi )$ surface and the plane at the bottom of the plot indicate the magnitude of the displacement angle whose gauge is shown on the right-hand side of (ix). There are two parameter regions with high fidelity with ${\delta}_{\mathrm{max}} (J; \tilde{\omega}, \phi ) \gtrsim 0.4\pi$ for $J=100$-$150$, while ${\delta}_{\mathrm{max}} (J; \tilde{\omega}, \phi )$ decreases to be ${\delta}_{\mathrm{max}} (J; \tilde{\omega}, \phi ) \sim 0$ in the region with smaller $\tilde{\omega}$ for $J=174.5$-$250$. }
	\label{fig:maxfidelity} 
\end{figure*} 
\begin{figure*} 
	\centering \includegraphics[bb = 0 0 415 357, clip, scale = 0.44]{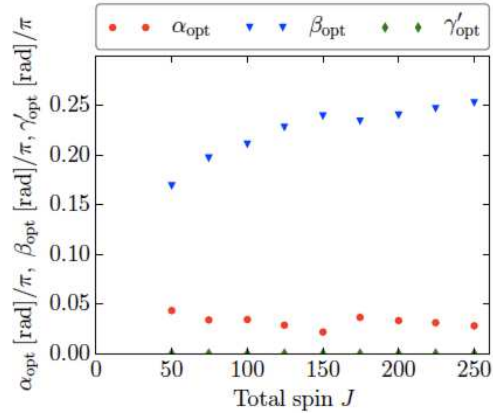}
	\caption{(Color Online) Plots of $J$ dependence of the angles ${\alpha}_{\mathrm{opt}}$, ${\beta}_{\mathrm{opt}}$, and ${\gamma}^{\prime}_{\mathrm{opt}}$, which are represented by the red dots, the blue triangles, and the green thin diamonds, respectively. The relative phase ${\gamma}^{\prime}_{\mathrm{opt}} = 0$ for all $J$. } 
	\label{fig:angles_vs_J}
\end{figure*} 

\section{\label{sec:s2}MSS generation via the Hamiltonian in Eq.~(1) and parameter optimization} 
\subsection{\label{sec:s2-1}Time dependence of fidelity, relative phase, and displacement angle} 
Starting from the initial state $|\Psi (J; \tilde{\omega} , \phi ; \tau =0) \rangle = | {\Phi}_{\mathrm{CSS}} (J; 0, \frac{\pi}{2} ) \rangle$, the state $|\Psi (J; \tilde{\omega} , \phi ; \tau) \rangle$ evolves under the Hamiltonian in Eq.~(1) and an MSS is formed. 
We plot the fidelity $F (J; \tilde{\omega} ,\phi ;\tau )$, the relative phase ${\gamma}_0^{\prime}$, and the displacement angle ${\delta}_0 ({\alpha}_0 ,{\beta}_0 )$, which are respectively defined in Eqs.~(5), (6), and (4), as functions of the rescaled evolution time $\tau$ for $J=50$, $J=74.5$, and $J=200$ in Figs.~\ref{fig:timedependence}. 
Here, in order to obtain $F (J; \tilde{\omega} ,\phi ;\tau )$,  ${\gamma}_0^{\prime}$, and ${\delta}_0 ({\alpha}_0 ,{\beta}_0 )$, the probability $| \langle {\Phi}_{\mathrm{MSS}} (J; \alpha ,\beta ,\gamma )| \Psi (J; \tilde{\omega} ,\phi ;\tau ) \rangle |^2$ is numerically maximized with respect to $\alpha$, $\beta$, and ${\gamma}^{\prime}$ by the basin-hopping method~\cite{basin}, which finds the global minimum or maximum of a smooth scalar function with one or more variables~\cite{Wales}. 
The first local maximum of the fidelity $F_{\mathrm{max}} (J; \tilde{\omega} ,\phi ) \equiv F (J; \tilde{\omega} ,\phi ; {\tau}_{\mathrm{max}})$ and its corresponding evolution time ${\tau}_{\mathrm{max}}$ are obtained from $F (J; \tilde{\omega} ,\phi ;\tau )$ in Figs.~\ref{fig:timedependence} by the brute-force search that starts from $\tau =0$ in the temporal order. 
The obtained ${\tau}_{\mathrm{max}}$'s are indicated by the black and thin dashed lines in Figs.~\ref{fig:timedependence}. 

In order to visually display the MSS creation, we plot the Q-functions of $|\Psi (J; \tilde{\omega} , \phi ; \tau ) \rangle$ at $\tau$'s indicated by the black and thin dotted lines (a)-(e) of Figs.~\ref{fig:timedependence} in Figs.~\ref{fig:QfuncMSS}. 
As shown in Figs.~\ref{fig:timedependence}, at the beginning of the time evolution, the fidelity $F (J; \tilde{\omega} ,\phi ;\tau )$ decreases, while the displacement angle ${\delta}_0 ({\alpha}_0 , {\beta}_0)$ increases. 
In this process, the state $|\Psi (J; \tilde{\omega} , \phi ; \tau ) \rangle$ is squeezed, which are illustrated in the Q-functions in Figs.~\ref{fig:QfuncMSS} (b). 
After that, the Q-function on the Bloch sphere is bent and tore off at $\alpha = 0$ and $\beta = \frac{\pi}{2}$ as we can see in Figs~\ref{fig:QfuncMSS} (b)-(c), and the two peaks of the Q-function move in the opposite directions as shown in Figs.~\ref{fig:QfuncMSS} (c)-(e). 
For $J = 74.5$ and $J = 200$, finite portions of the probability distribution remain around $\alpha = 0$ and $\beta = \frac{\pi}{2}$, which causes a decrease in the fidelity $F_{\mathrm{max}} (J; \tilde{\omega} ,\phi )$ of the first local maximum. 

\subsection{\label{sec:s2-2}Driving frequency and phase dependence of fidelity and displacement angle} 
We optimize the frequency $\tilde{\omega}$ and the phase $\phi$ of the driving field with respect to the fidelity $F_{\mathrm{max}} (J; \tilde{\omega} , \phi )$ and the displacement angle ${\delta}_{\mathrm{max}} (J; \tilde{\omega}, \phi )$ and obtain the $J$ dependences of the optimized fidelity $F_{\mathrm{opt}} (J)$, the displacement angle ${\delta}_{\mathrm{opt}} (J)$, their corresponding driving-field parameters ${\tilde{\omega}}_{\mathrm{opt}}$ and ${\phi}_{\mathrm{opt}}$, and the evolution time ${\tau}_{\mathrm{opt}}$. 
Here, the displacement angle ${\delta}_{\mathrm{opt}}$ is calculated from ${\alpha}_{\mathrm{opt}} (J)$ and ${\beta}_{\mathrm{opt}} (J)$. 
We plot $F_{\mathrm{max}} (J; \tilde{\omega} , \phi )$ and ${\delta}_{\mathrm{max}} (J; \tilde{\omega}, \phi )$ as functions of $\tilde{\omega}$ and $\phi$ in Figs.~\ref{fig:maxfidelity} for $J=50$-$250$ and obtain ${\tilde{\omega}}_{\mathrm{opt}}$ and ${\phi}_{\mathrm{opt}}$ by the brute-force search such that $F_{\mathrm{opt}} (J)$ is the maximum of $F_{\mathrm{max}} (J; \tilde{\omega} , \phi )$ with respect to $\tilde{\omega}$ and $\phi$ in the parameter region satisfying ${\delta}_{\mathrm{opt}} (J) > 0.4 \pi$. 
Figures~\ref{fig:maxfidelity} are plotted against $51\times 51$ pairs of $\tilde{\omega}$ and $\phi$ and we do finer calculation with the precision of $\Delta \tilde{\omega} = \pi \times 10^{-4}$ and $\Delta \tilde{\phi} = \pi \times 10^{-4}$ around the peaks obtained from Figs.~\ref{fig:maxfidelity} in order to estimate $\tilde{\omega}$ and $\phi$ for $J \geq 150$. 
The $J$ dependence of ${\alpha}_{\mathrm{opt}}$, ${\beta}_{\mathrm{opt}}$, and ${\gamma}^{\prime}_{\mathrm{opt}}$ are shown in Fig.~\ref{fig:angles_vs_J}. 
The plot of ${\gamma}_{\mathrm{opt}}^{\prime}$ indicates that the MSS creation via the Hamiltonian in Eq.~(1) is robust against the fluctuations in the spin number, the driving-field frequency, and the evolution time. 

\section{\label{sec:s3}Interferometry using MSSs} 
\subsection{\label{sec:s3-1}Idealistic case} 
Suppose we can prepare a perfect MSS given by Eq.~(2). 
The nonclassicality of the MSS can be observed by the following procedure~\cite{Bollinger1}: 
First, let an MSS $|{\Phi}_{\mathrm{MSS}} (J; \alpha ,\beta ,\gamma ) \rangle $ rotate about the $z$ axis by a small angle $\theta$, which results in the state $|{\Phi}_{\mathrm{MSS}} (J; \alpha ,\beta ,\gamma ) {\rangle}_{\theta}$, i.e., 
\begin{equation} 
	|{\Phi}_{\mathrm{MSS}} (J; \alpha ,\beta ,\gamma ) {\rangle}_{\theta} = e^{-i{\hat{J}}_z \theta} |{\Phi}_{\mathrm{MSS}} (J; \alpha ,\beta ,\gamma ) \rangle . \label{eq:Phi-MSS-theta}
\end{equation} 
Then, measure the parity of ${\hat{\sigma}}_x$ of the spin ensemble: 
\begin{widetext} 
\begin{align} 
	_{\theta} {\langle} {\Phi}_{\mathrm{MSS}} (J; \alpha ,\beta ,\gamma ) |  {\hat{\sigma}}_x^{\otimes N} | {\Phi}_{\mathrm{MSS}} (J; \alpha ,\beta ,\gamma ) {\rangle}_{\theta} 
	=& \frac{1}{A^2(J; \alpha ,\beta ,\gamma )} \biggl \{ 
		2 \cos {[2J(\theta - \alpha )]} \ {\cos}^{2J} (\theta + \alpha ) {\sin}^{2J} \beta \nonumber \\ 
	&+ \sum_{n=0}^{2J} \ _{2J}{\mathrm{C}}_n {\cos}^{2(2J - n)} \frac{\beta}{2} \ {\sin}^{2n} \frac{\beta}{2} \ \left [ e^{i (2J\theta + {\gamma}^{\prime} )} e^{-2im\theta} + \mathrm{h.c.} \right ]
	\biggr \} \nonumber \\ 
	\simeq & \frac{1}{2} \sum_{n=0}^{2J} \ _{2J}{\mathrm{C}}_n {\cos}^{2(2J - n)} \frac{\beta}{2} \ {\sin}^{2n} \frac{\beta}{2} \ \left [ e^{i (2J\theta + {\gamma}^{\prime} )} e^{-2im\theta} + \mathrm{h.c.} \right ], \label{eq:expect-parity-x-intermediate} 
\end{align} 
\end{widetext} 
where $A (J; \alpha ,\beta ,\gamma )$ is given in Eq.~(\ref{eq:norm}) and we neglect the terms proportional to ${\sin}^{2J} \beta$ on the right-hand side of the last equality, since it is as small as $\sim O (10^{-23})$ in the parameter region of $N = 2J \sim O (10^2)$ and $\beta \sim 0.2 \pi$ that $|{\Psi}_{\mathrm{opt}} (J) \rangle$ satisfies for $J=50$-$250$ as we plot in Fig.~\ref{fig:angles_vs_J}. 
The parameter region also verifies another important approximation: 
The sum on the right-hand side of Eq.~(\ref{eq:expect-parity-x-intermediate}) can be well approximated by the Gaussian integral, since the term $_{2J}{\mathrm{C}}_n {\cos}^{2(2J - n)} \frac{\beta}{2} \ {\sin}^{2n} \frac{\beta}{2}$ in Eq.~(\ref{eq:expect-parity-x-intermediate}) can be considered as the binomial distribution of the number of success in a sequence of $N$ independent trials with the success rate of ${\sin}^2 \frac{\beta}{2}$ and the absolute value of its skewness is approximately given by 
\begin{equation} 
	\frac{|1 - 2 {\sin}^2 \frac{\beta}{2} |}{\sqrt{\frac{1}{2} {\sin}^2 \beta}} \sim 0.3 < \frac{1}{3},   
\end{equation} 
which indicates this binomial distribution can be well approximated by the normal distribution. 
Therefore, the expectation value of the parity of ${\hat{\sigma}}_x$ is approximately obtained as 
\begin{align} 
	&_{\theta} {\langle} {\Phi}_{\mathrm{MSS}} (J; \alpha ,\beta ,\gamma ) |  {\hat{\sigma}}_x^{\otimes N} | {\Phi}_{\mathrm{MSS}} (J; \alpha ,\beta ,\gamma ) {\rangle}_{\theta} \nonumber \\ 
	\simeq & \frac{1}{2} \int_{-\infty}^{\infty} \frac{dx}{\sqrt{\pi J {\sin}^2 \beta}} \biggl \{ \exp \biggl [ - \frac{(x - 2 J {\sin}^2 \frac{\beta}{2})^2 }{J {\sin}^2 \beta } \nonumber \\ 
	& \hspace{3.5cm} - 2 i \left ( \theta x - J \theta - \frac{{\gamma}^{\prime}}{2} \right ) \biggr ]  + \mathrm{h.c.} \biggr \} \nonumber \\ 
	=& e^{- J {\theta}^2 {\sin}^2 \beta} \cos {\left (2 J \theta \cos {\beta} + {\gamma}^{\prime} \right )}, \label{eq:expect-parity-x}
\end{align} 
and the variance of the parity of ${\hat{\sigma}}_x$ is given by 
\begin{align} 
	\langle (\Delta {\hat{\sigma}}_x^{\otimes N})^2 \rangle = 1 - e^{- 2 J {\theta}^2 {\sin}^2 \beta} {\cos}^2 \left (2 J \theta \cos {\beta} + {\gamma}^{\prime} \right ). \label{eq:variance-parity-x}
\end{align} 
Equation~(\ref{eq:variance-parity-x}) implies that one can expect to observe the fringe for the rotation angle $\theta$ satisfying $| \theta | \lesssim ( 2J {\sin}^2 \beta )^{-1/2}$. 
This range of the rotation angle allows us to observe about $\sqrt{2J} \ \frac{{\cos}^2 \beta }{\pi \sin \beta } \sim 0.35 \times \sqrt{2J}$ fringes for $\beta \sim 0.2 \pi$, which implies that we can expect to observe four fringes for a $J=74.5$ spin ensemble and seven fringes for a $J=200$ spin ensemble if a perfect MSS can be prepared. 
We also note that the width of the single fringe $\Delta \theta$ is given by 
\begin{equation} 
	\Delta \theta = \frac{\pi}{2J {\cos}^2 \beta} \sim 0.76\pi \times J^{-1} [\mathrm{rad} ] \label{eq:precision}
\end{equation} 
for $\beta \sim 0.2 \pi$, which implies that an MSS can be utilized as a probe of Heisenberg-limited spectroscopy. 
On the other hand, when the state is mixed, i.e., 
\begin{align} 
	&{\hat{\rho}}_{\mathrm{mix}} (J; \alpha ,\beta ) \nonumber \\ 
	=& \frac{1}{A^2(J;\alpha ,\beta ,\gamma )} ( |{\Phi}_{\mathrm{CSS}} (J; \alpha ,\beta ) \rangle \langle {\Phi}_{\mathrm{CSS}} (J; \alpha ,\beta ) | \nonumber \\ 
	& + |{\Phi}_{\mathrm{CSS}} (J; - \alpha ,\pi - \beta ) \rangle \langle {\Phi}_{\mathrm{CSS}} (J; - \alpha ,\pi - \beta ) | ),  
\end{align} 
the variance of the parity of ${\hat{\sigma}}_x$ after the rotation about the $z$ axis by an angle $\theta$ is given by 
\begin{align} 
	\langle (\Delta {\hat{\sigma}}_x^{\otimes N})^2 \rangle &= 1 - \frac{[{\cos}^{2J} (\theta + \alpha ) + {\cos}^{2J} (\theta - \alpha ) ] {\sin}^{2J} \beta }{2(1 + {\cos}^{2J} \alpha \ {\sin}^{2J} \beta \cos {{\gamma}^{\prime}} )} \nonumber \\ 
	& \simeq 1,  
\end{align} 
and no fringes can be observed. 

\subsection{\label{sec:s3-2}MSSs with spin number fluctuations} 
As shown in Eq.~(\ref{eq:variance-parity-x}) in the previous subsection, a perfect MSS manifests fringes of $\langle (\Delta {\hat{\sigma}}_x^{\otimes N})^2 \rangle$ whose width is given by the Heisenberg-limit scaling law $\propto J^{-1}$. 
The fringes generated by $|{\Psi}_{\mathrm{opt}} (J) \rangle$; however, are expected to be degraded by the imperfection of $|{\Psi}_{\mathrm{opt}} (J) \rangle$. 
Moreover, the number of spins in an ensemble may well have finite fluctuation in experiments, for instance, the number of atoms is fluctuating as $380 \pm 15$ in Ref.~\cite{Strobel}, and the fringes may fade away, depending on the magnitude of the number fluctuations. 
In order to investigate the robustness of the fringes generated by $|{\Psi}_{\mathrm{opt}} (J) \rangle$ against the imperfection of the fidelity to the perfect MSS $|{\Phi}_{\mathrm{MSS}} (J; {\alpha}_{\mathrm{opt}} ,{\beta}_{\mathrm{opt}}, {\gamma}_{\mathrm{opt}} ) \rangle$ and the spin-number fluctuation, we numerically calculate the fringes of $\langle (\Delta {\hat{\sigma}}_x^{\otimes N})^2 \rangle$ generated by $|{\Psi}_{\mathrm{opt}} (J) \rangle$ whose spin number is Gaussian-fluctuating, i.e., the probability to have $N$ spins can be expressed as the normal distribution $P ({\bar{N}}, \sigma ;N)$ with the mean value of $\bar{N}$ and the standard deviation $\sigma$ given by 
\begin{equation} 
	P ({\bar{N}}, \sigma ;N) = \frac{1}{\sqrt{2\pi {\sigma}^2}} e^{-\frac{(N - \bar{N} )^2}{2{\sigma}^2}}. \label{eq:probN} 
\end{equation} 
In the case of Ref.~\cite{Strobel}, the mean and the standard deviation of the spin number are given by $\bar{N} = 380$ and $\sigma /J < 3.9\%$, respectively. 
In our calculation of $\langle (\Delta {\hat{\sigma}}_x^{\otimes N})^2 \rangle$ with finite spin-number fluctuation, 250 pseudo-random spin numbers with the probability density given by Eq.~(\ref{eq:probN}) are generated by the Mersenne Twister method so that the difference between the mean spin number of $N_{\mathrm{trial}} = 250$ traials and $\bar{N}$ in Eq.~(\ref{eq:probN}) and their respective standard deviations ${\sigma}_{\mathrm{trial}}$ and $\sigma$ satisfy $| N_{\mathrm{trial}} - \bar{N} | \leq 0.01 \times \bar{N}$ and $|{\sigma}_{\mathrm{trial}} - \sigma | \leq 0.1 \times \sigma$.  
\begin{figure*} 
	\centering \includegraphics[bb = 0 0 1045 767, clip, scale = 0.44]{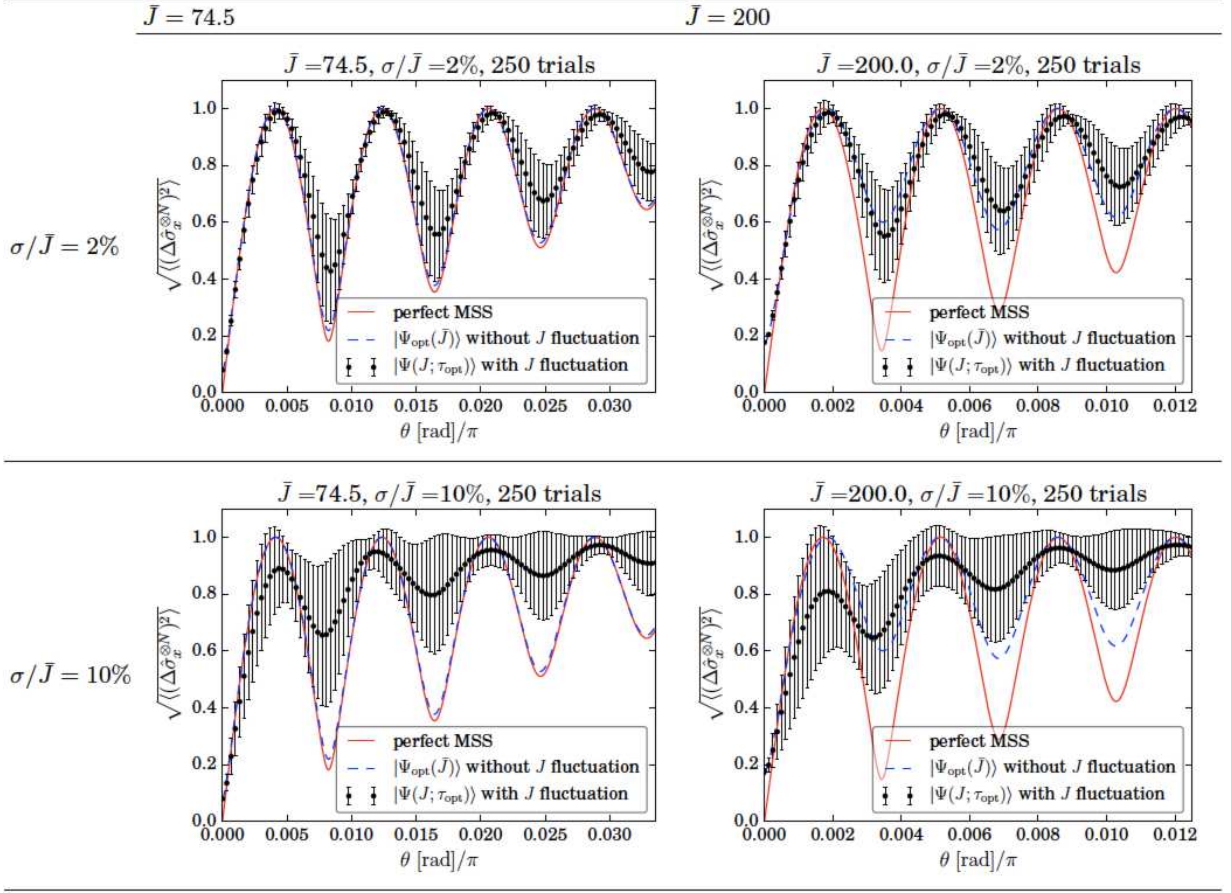}
	\caption{(Color Online) Fringes of $\sqrt{\langle (\Delta {\hat{\sigma}}_x^{\otimes N} )^2 \rangle}$ generated by $|{\Psi}_{\mathrm{opt}} (J) \rangle$ with the spin-number fluctuations of $2\%$ and $10\%$ for $\bar{J} =74.5$ and $\bar{J} =200$. The red solid curves and the blue dashed curves express the fringes produced by the perfect MSS $|{\Phi}_{\mathrm{MSS}} (J; {\alpha}_{\mathrm{opt}} ,{\beta}_{\mathrm{opt}}, {\gamma}_{\mathrm{opt}} ) \rangle$ and $|{\Psi}_{\mathrm{opt}} (J) \rangle$ without spin-number fluctuation. The black dots with the black-solid error bars respectively represent the mean values and the standard deviations of 250 trials of fringe experiments, where the spin-number fluctuation is given by Eq.~(\ref{eq:probN}) and $|\Psi (J; {\tau}_{\mathrm{opt}} ) \rangle$ is prepared after the optimised evolution time ${\tau}_{\mathrm{opt}}$ for $\bar{J}$. } 
	\label{fig:fringevarN} 
\end{figure*} 

We show the fringes for $\bar{J} = 74.5$ and $\bar{J} = 200$ without and with the spin-number fluctuations of $2\%$ and $10\%$ in Figs.~\ref{fig:fringevarN}. 
The fringes for the perfect MSS and $|{\Psi}_{\mathrm{opt}} (\bar{J} ) \rangle$ without the spin-number fluctuation almost coincide with each other in the case of $\bar{J} = 74.5$, when the fidelity to the perfect MSS exceeds 0.99. 
On the other hand, the magnitudes of the fringes generated by $|{\Psi}_{\mathrm{opt}} (\bar{J} ) \rangle$ decrease in comparison with the perfect MSS even without the spin-number fluctuation in the case of $\bar{J} = 200$, when the fidelity to the perfect MSS degraded to be $\sim 0.86$; however, the magnitude of the fringe created by $|{\Psi}_{\mathrm{opt}} (\bar{J} ) \rangle$ is diminished more slowly than the perfect MSS with respect to the rotation angle $\theta$ and the positions of the fringe peaks does not change from those of the perfect MSS. 
Thus we can expect to observe the fringes and make use of it to estimate the rotation angle up to the spin number of $N = 500$ at least when the number of spins can be deterministically prepared. 
Figure~\ref{fig:fringevarN} also imply the robustness against the spin-number fluctuation of $\lesssim 10\%$. 
The fringes with the spin number fluctuation of $10\%$ is not suitable for the rotation-angle measurement; however, they still manifest the nonclassicality, since we can clearly see the region $\sqrt{\langle (\Delta {\hat{\sigma}}_x^{\otimes N})^2 \rangle} < 1$ on either side of the first peak of $\sqrt{\langle (\Delta {\hat{\sigma}}_x^{\otimes N})^2 \rangle}$ at $\theta \sim 0.38 \pi \times J^{-1}$ given in Eq.~(\ref{eq:precision}) as shown in Figs.~\ref{fig:fringevarN}. 

\begin{figure*} 
	\centering \includegraphics[bb = 0 0 1046 769, clip, scale = 0.44]{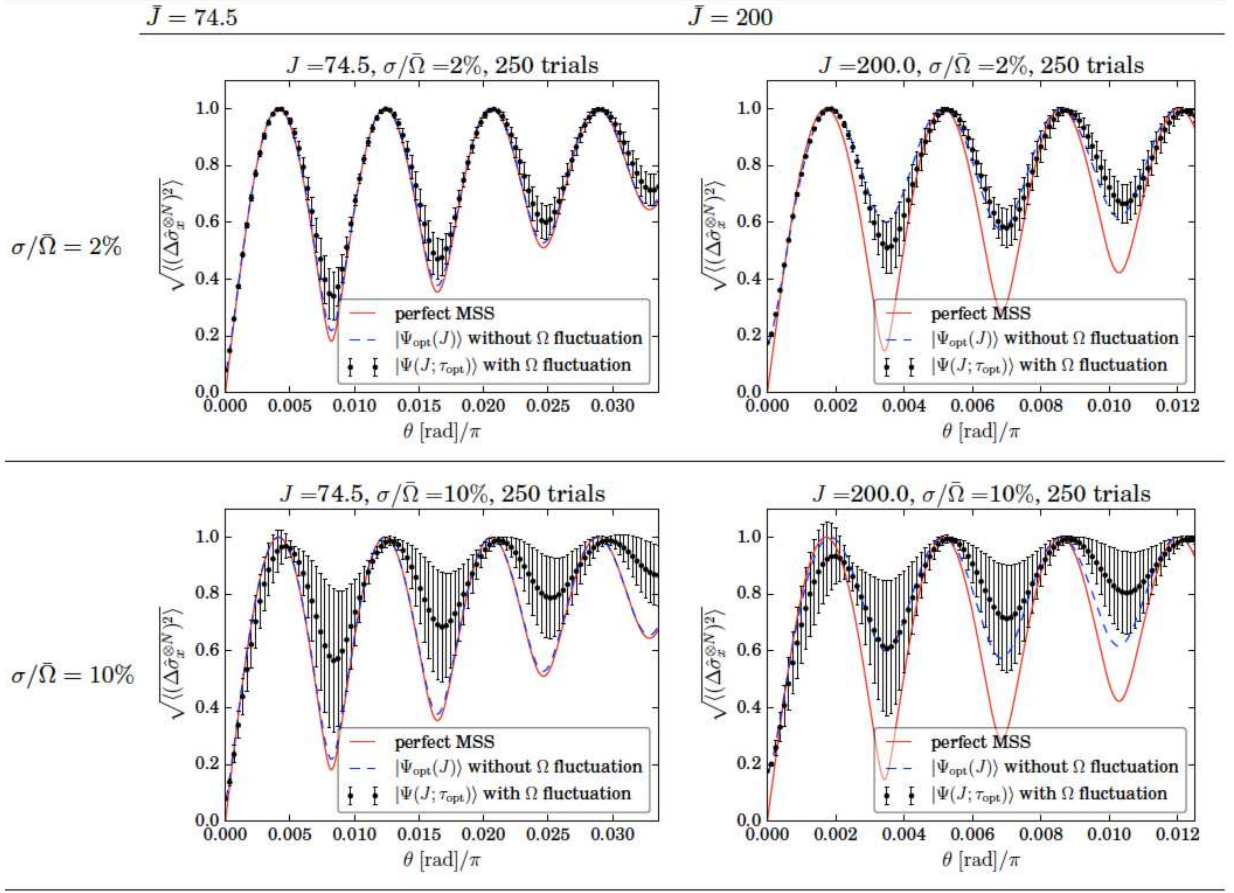}
	\caption{(Color Online) Fringes of $\sqrt{\langle (\Delta {\hat{\sigma}}_x^{\otimes N})^2 \rangle}$ generated by $|{\Psi}_{\mathrm{opt}} (J) \rangle$ with the fluctuation in the driving-field magnitude of $2\%$ and $10\%$ for $\bar{J} =74.5$ and $\bar{J} =200$. The red solid curves and the blue dashed curves express the fringes produced by the perfect MSS $|{\Phi}_{\mathrm{MSS}} (J; {\alpha}_{\mathrm{opt}} ,{\beta}_{\mathrm{opt}}, {\gamma}_{\mathrm{opt}} ) \rangle$ and $|{\Psi}_{\mathrm{opt}} (J) \rangle$ without the fluctuation in $\Omega$, i.e., $\Omega = \bar{\Omega}$. The black dots with the black-solid error bars respectively represent the mean values and the standard deviations of 250 trials of fringe experiments with the $\Omega$ distributed randomly between $(1 \pm \sigma ) \bar{\Omega}$. } 
	\label{fig:fringevarOmega}  
\end{figure*} 
\begin{figure*} 
	\centering \includegraphics[bb = 0 0 1043 1136, clip, scale = 0.44]{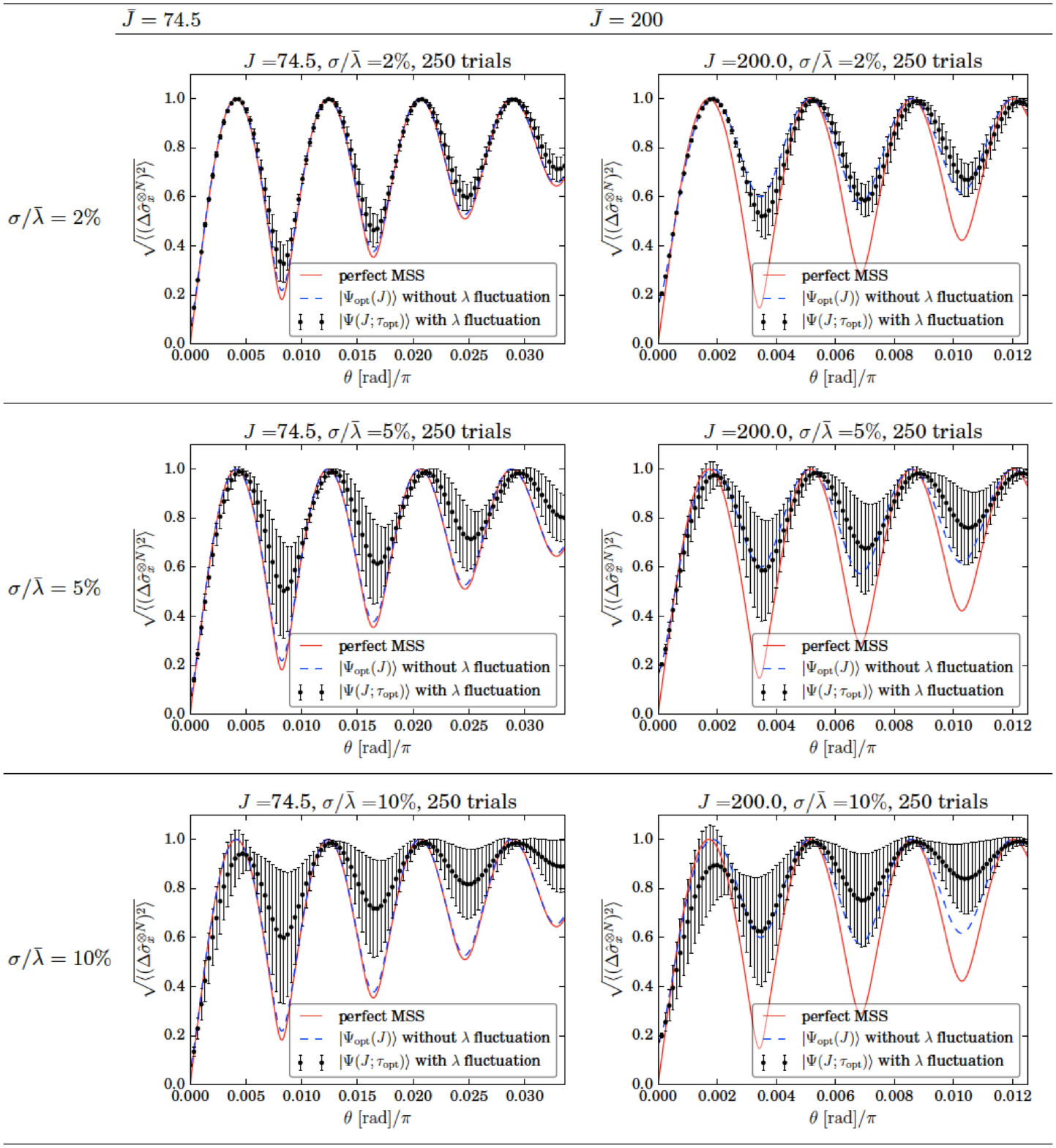}
	\caption{(Color Online) Fringes of $\sqrt{\langle (\Delta {\hat{\sigma}}_x^{\otimes N})^2 \rangle}$ generated by $|{\Psi}_{\mathrm{opt}} (J) \rangle$ with the fluctuation in the nonlinear interaction energy $\lambda$ of $2\%$, $5\%$, and $10\%$ for $\bar{J} =74.5$ and $\bar{J} =200$. The red solid curves and the blue dashed curves express the fringes produced by the perfect MSS $|{\Phi}_{\mathrm{MSS}} (J; {\alpha}_{\mathrm{opt}} ,{\beta}_{\mathrm{opt}}, {\gamma}_{\mathrm{opt}} ) \rangle$ and $|{\Psi}_{\mathrm{opt}} (J) \rangle$ without the fluctuation in $\lambda$, i.e., $\lambda = \bar{\lambda}$. The black dots with the black-solid error bars respectively represent the mean values and the standard deviations of 250 trials of fringe experiments with the $\lambda$ distributed randomly between $(1 \pm \sigma ) \bar{\lambda}$. } 
	\label{fig:fringevarLambda} 
\end{figure*} 

\subsection{\label{sec:s3-3}Other noise sources} 
The other major noise sources are the fluctuations in the magnitude of the driving field $\Omega$ and the evolution time $t_{\mathrm{opt}}$ of an MSS creation during a series of trials to obtain fringes. 
Here, $t_{\mathrm{opt}}$ is well controllable to within the order of $\sim \mu \mathrm{s}$ as well as driving-field parameters $\tilde{\omega}$ and $\phi$ whose fluctuations are negligible when the interaction strength is given by $\sim [\mathrm{Hz}]$; however, it can be a major source of fluctuations when the achievable interaction strength gets larger to be $\sim [\mathrm{kHz}]$. 

The fluctuation in $\Omega$ can be caused by the fluctuation in the energy splitting between two internal degrees of freedom comprising a pseudo spin. 
We assume that $\Omega$ uniformly distributes between $[(1- \sigma ) \Omega, (1+\sigma ) \Omega]$ and simulate the fringes produced by $|{\Psi}_{\mathrm{opt}} (\bar{J} ) \rangle$ with the fluctuation in $\Omega$ of $2\%$ and $10\%$ for $J = 74.5$ and $J=200$ as shown in Figs.~\ref{fig:fringevarOmega}. 
Here, we generate 250 pseudo-random value of $\Omega$'s, each of which are un-correlated, and ensure that the average of $\Omega$ of the 250 trials, ${\Omega}_{\mathrm{trial}}$, satisfies $|{\Omega}_{\mathrm{trial}} - \bar{\Omega} | < 0.01 \times \bar{\Omega}$. 
We can observe the fringes even when $\Omega$ fluctuates $10\%$ of its mean value. 

The fluctuation in $t_{\mathrm{opt}}$ is equivalent to the fluctuation in the nonlinear interaction energy $\lambda$. 
So, we assume that $\lambda$ has a uniform distribution between $[(1 - \sigma ) \lambda , (1 + \sigma ) \lambda ]$ and obtain the fringes of $\sqrt{\langle (\Delta {\hat{\sigma}}_x^{\otimes N})^2 \rangle}$ produced by the MSSs with random $\lambda$ for $N = 149$ and $N = 400$ and for $\sigma / \lambda = 2\%$, $5\%$, and $10\%$. 
As in the case of fluctuating $\Omega$, we can expect to observe interference fringes when $\lambda$ or $t_{\mathrm{opt}}$ fluctuates $10\%$ of its magnitude. 

 
\end{document}